\newcommand{\commentt}[2]{#1}
\newcommand{\bff}{}
\newcommand{\comment}[1]{}

\newcommand{\tr}{{\rm tr}}
\newcommand{\cov}{{\rm Cov}}
\newcommand{\E}{{\rm E}}

\newcommand{\X}{{\bff X}}
\newcommand{\var}{{\rm Var}}
\newcommand{\Y}{{\bff Y}}
\newcommand{\Z}{{\bff Z}}

\newcommand{\A}{{\bff A}}
\newcommand{\B}{{\bff B}}
\newcommand{\I}{{\bff I}}
\newcommand{\M}{{\bff M}}
\newcommand{\V}{{\bff V}}
\newcommand{\U}{{\bff U}}

\newcommand{\e}{{\bff e}}
\newcommand{\x}{{\bff x}}
\newcommand{\vv}{{\bff v}}
\newcommand{\f}{{\bff f}}
\newcommand{\Pf}{{\bff P}}
\commentt{
\documentclass[a4paper,11pt]{article}
\usepackage{graphicx,slashbox,amsthm}
\usepackage[full]{harvard}

\usepackage[ps,dvips]{xy}
\title{Efficient simulation of high dimensional Gaussian vectors}
\author{Nabil Kahal\'e
\thanks{\emph{ESCP Europe, Labex Refi and Big Data Research Center, 
75011 Paris,
France; e-mail: nkahale@escpeurope.eu.}}
}
\date{\today}
\usepackage{amstext}
\usepackage{amssymb}
\usepackage{amsmath}
\usepackage{setspace}
\usepackage[margin=1in]{geometry}
\usepackage[english]{babel}
\commentt{\bibliographystyle{dcu}}{}
\usepackage{color}
\usepackage{hyperref}
\begin{document}

\newtheorem{example}{Example}[section]
\newtheorem{theorem}{Theorem}[section]
\newtheorem{conjecture}{Conjecture}[section]
\newtheorem{lemma}{Lemma}[section]
\newtheorem{proposition}{Proposition}[section]
\newtheorem{remark}{Remark}[section]
\newtheorem{corollary}{Corollary}[section]
\newtheorem{definition}{Definition}[section]
\numberwithin{equation}{section}
\maketitle
\newcommand{\ABSTRACT}[1]{\begin{abstract}#1\end{abstract}}
\newcommand{\citep}{\cite}
}
{
\documentclass[moor]{informs3}
\usepackage{natbib}

\newtheorem{lemma}{Lemma}
\newtheorem{theorem}{Theorem}
\newtheorem{definition}{Definition}

 \NatBibNumeric
 \def\bibfont{\small}%
 \def\bibsep{\smallskipamount}%
 \def\bibhang{24pt}%
 \def\BIBand{and}%
 \def\newblock{\ }%
 \bibpunct[, ]{[}{]}{,}{n}{}{,}%
              
\begin{document}
\RUNAUTHOR{Kahale}
\RUNTITLE{Efficient Simulation of Gaussian Vectors}
\TITLE{Efficient simulation of high dimensional Gaussian vectors}
\ARTICLEAUTHORS{
\AUTHOR{Nabil Kahale}
\AFF{ESCP Europe, Labex R\'efi and Big Data Research Center, 75011 Paris, France, \EMAIL{nkahale@escpeurope.eu} \URL{}}
}
}
\ABSTRACT{%
We describe a Markov chain Monte Carlo  method to approximately simulate a centered  \(d\)-dimensional Gaussian vector \(X\)   with given  covariance matrix. The standard Monte Carlo method is based on the Cholesky decomposition, which takes cubic time  and has  quadratic storage cost  in  \(d\).  In contrast, the storage cost of our algorithm is linear in \(d\).    We give a bound on the quadractic Wasserstein distance between the distribution of our sample and the target distribution. Our method can be used to estimate the expectation of \(h(X)\), where \(h\) is a real-valued function of \(d\) variables. Under certain conditions, we show that the mean square error of our method is inversely proportional to its running time.  We also prove that, under suitable conditions, our method is faster than  the standard Monte Carlo method  by a factor nearly proportional to \(d\). A numerical example  is given.
}
\commentt{
}
{
\bibliographystyle{informs2014} 
\maketitle
}
\section{Introduction}
Monte Carlo simulation of  Gaussian vectors  is commonly used in  a  variety of fields such
as
weather prediction~\citep{gel2004}, finance~\cite[Chap. 13]{Hull12}, and machine learning~\cite{russo2014learning,russoBVR2016}.
This paper considers the problem of efficiently sampling a  \(d\)-dimensional
Gaussian vector
 \(\X\)  with  a given mean and a given \(d\times d\)  covariance matrix \(\V\). Since any Gaussian random variable is an affine function of a standard Gaussian random variable,   we assume throughout the paper  that the components of \(\X\) are standard Gaussian random variables, and so the diagonal elements of \(\V\)  are \(1\). Then \(\X\) can be simulated \cite[Subsection 2.3.3]{glasserman2004Monte} as follows. Let \(\Z\)  be a \(d\)-dimensional vector of independent standard Gaussian random variables, and  let \(\A\) be  a \(d\times d\) matrix such that \begin{equation}\label{eq:aat}
\A\A^{T}=\V.
\end{equation} Then \(\A\Z\sim N(\bff {\bff 0},\V)\), i.e. \(\A\Z\) is a \(d\)-dimensional Gaussian vector with covariance matrix \(\V\). 

Such a matrix  \(\A\) \comment{is called a Cholesky factorization if  \(\A\) is lower triangular. If \(\V\) is positive definite, a Cholesky factorization exists and}can be computed in \(O(d^{{3}})\) time and \(O(d^{{2}})\) space using Cholesky factorization or one of its variants~\cite[Subsections 4.2.5 and 4.2.8]{golub2013matrix}.\comment{ \cite[Subsection 2.3.3]{glasserman2004Monte}.If \(\V\) is positive semi-definite,
 a matrix \(\A\) satisfying~\eqref{eq:aat} can also be calculated in \(O(d^{^{3}})\) time
and \(O(d^{^{2}})\) storage cost  using a variant of the Cholesky factorization~\cite[Subsection 4.2.8]{golub2013matrix}.}  Once \(\A\) is calculated,  \(\A\Z\) can be computed in \(O(d^{2})\) time. But, in several applications (see e.g.~\cite{gel2004}),   \(d\) is in the tens of thousands or more,  and so
the calculation of  a Cholesky factorization on a standard   computer may not be possible in practice,  due to the high running time and/or storage cost.  
Alternative methods for generating  Gaussian vectors have been developed
in special cases.  For instance, exact
and efficient simulation of   Gaussian processes on a regular grid
 in \(\mathbb{R}^{q}\), \(q\geq1\),   can be performed~\comment{ a symmetric matrix \(\A\) satisfying~\ref{eq:aat}
can be approximately calculated~\cite{waugh1967fractional} via a Taylor series.
Such a matrix \(\A\) can be used to generate a centered Gaussian vector whose covariance
matrix is close to \(\V\).}\citep{WoodChan94,dietrich1997fast} using Fast Fourier transforms    if the covariance matrix is         stationary with respect
 to translations.  
  Similar Fast  Fourier Transform methods can be used for exact
  simulation of   fractional Brownian surfaces on a regular mesh~\citep{stein2002fast}.
Sparse Cholesky decomposition~\citep{rue2001fast} and iterative methods~\citep{aune2013iterative} have been proposed  to generate
 efficiently Gaussian vectors when the precision matrix \(\V^{-1}\) is  sparse. 
\comment{for drawing a Gaussian sample from the positive orthant is
  analysed in~\cite{kannan1996sampling} .}   

This paper develops a new Markov Chain Monte Carlo method for approximate generation of a Gaussian vector \(\X\) with
correlation matrix \(\V\). Our method is straightforward to implement and can be applied to any correlation matrix
\(\V\) whose elements  are known or  easy to
compute. It has a total storage cost of \(O(d)\). At iteration \(n\), it
produces a \(d\)-dimensional vector  \(\X_{n}\) whose distribution converges
(according to the quadratic Wasserstein distance) to   \(N({\bff 0},\V)\) as \(n\)
goes to infinity. Assuming each element
of \(\V\) can be computed in \(O(1)\) time, the running time of each iteration is  \(O(d)\).  Our method can for instance be used  to approximately simulate  spatial Gaussian processes of various  types  such as  Mat\'ern, powered exponential, and spherical   on \emph{any subset} of size \(d\) of \(\mathbb{R}^{2}\) with \(O(d)\) storage cost (background on spatial statistics can be found in~\cite{diggle2003introduction}). While FFT methods can  simulate such processes on regular grids, certain applications (e.g.~\cite{gel2004}) require the simulation of  spatial Gaussian processes on  non-regular subsets of \(\mathbb{R}^{2}\). 

We now describe our method in more detail.  
 Let  \((i_{n}\)), \(n\geq0\), be a deterministic or a random  sequence
 in \(\{1,\ldots ,d\}\),  and let \((g_{n}\)), \(n\geq0\),
 be a sequence of independent standard Gaussian random variables, independent of \((i_{n})\), \(n\geq0\). Define the Markov chain of \(d\)-dimensional column vectors  \(\X_{n}\), \(n\ge0\),
as follows. Let  \(\X_{0}=\bff0\) and, for   \(n\geq0\),  let
 \begin{equation}\label{eq:xndefsimple}
\X_{n+1}=\X_{n}+(g_{n}-\e_{i_{n}}^{T}\X_{n})(\V \e_{i_{n}}),
\end{equation} where \(\e_{i}\) is the  \(d\)-dimensional  column vector whose
\(i\)-th coordinate is \(1\), and remaining coordinates are \(0 \) (if \(t\in\mathbb{R}\) and \(u\) is a vector, \(tu\) is the scalar product of \(t\) and \(u\)).   Since \(\V \e_{i_{n}}\) is the \(i_{n}\)-th column of \(\V\),   \(\X_{n+1}\)   
can be calculated from \(\X_{n}\) in \(O(d)\) time, with  storage cost \(O(d)\). The motivation behind~\eqref{eq:xndefsimple} is explained in Section~\ref{se:Notation}, where we also show that~\eqref{eq:xndefsimple} is a variant  of the hit-and-run algorithm. A general  description of the hit-and-run algorithm can be found in~\cite{smith1984}.

 Section~\ref{se:random} shows that, if  \(i_{n}\)  are independent random variables   uniformly distributed over \(\{1,\dots,d\}\), then   the quadratic Wasserstein distance between the distribution of \(\X_{n}\) and  \(N({\bff 0},\V)\)  is at most  \(d/\sqrt{n}\).   The quadratic Wasserstein distance between two probability distributions     \(\mu\) and \(\mu'\)  over \(\mathbb{R}^{d}\) is defined as \begin{equation}\label{eq:defWasserstein}
\mathcal{W}_{2}(\mu,\mu')=(\inf_{\Y\sim\mu,\Y'\sim\mu'}\E(||\Y-\Y'||^{2}))^{1/2}.
\end{equation}  To put this
result into perspective, denote by \(\mu_{\epsilon}\)  the distribution of \(N({\bff0},(1-\epsilon )\V)\), for \(0\leq\epsilon\leq1\). Then \(\mathcal{W}_{2}(\mu_{\epsilon},\mu_{0})=(1-\sqrt{1-\epsilon})\sqrt{d}\),
by~\cite[Eq. 16]{dowson1982frechet}. Thus, after \(n=O(d\epsilon^{-2})\) steps,
which can be performed in  \(O(d^{2}\epsilon^{-2})\) total time,
 the quadratic Wasserstein distance between the distribution of \(\X_{n}\) and  \(\mu_{0}\)  is at most  \(\mathcal{W}_{2}(\mu_{\epsilon},\mu_{0})\).  Section~\ref{se:Lipschitz} shows that, if  \(h\) is a real-valued function on \(\mathbb{R}^{d}\)  satisfying certain conditions, and  \(i_{0},\dots,i_{n-1}\)  are independent random variables   uniformly distributed over \(\{1,\dots,d\}\),
then   \(m=\E(h(\X))\), where  \(\X\sim N({\bff 0},\V)\),   is well approximated  by \(n^{-1}\sum ^{n-1}_{j=0}h(\X_{j})\). More precisely,
Theorem~\ref{th:MSE} give explicit bounds on the mean square error   \begin{equation*}
\text{MSE}(n)=\E((\frac{\sum ^{n-1}_{j=0}h(\X_{j})}{n}-m)^{2})
\end{equation*}   
of this estimator. For instance, if \(h\)  is \(\kappa\)-Lipschitz, 
Theorem~\ref{th:MSE} implies that \(n\text{MSE}(n)\leq18\kappa^{2}d^{2}\). We give an example with \(n=\Theta(d)\)  where this bound is tight, up to a constant.    To our knowledge,  for general \(V\), no previous methods achieve a similar tradeoff between the running time and the Wasserstein distance, or  between the running time and the mean square error, when \(n=\Theta(d)\). Section~\ref{se:asymptotics}
 assumes that \(\V\) is positive definite and shows that, under
suitable conditions, \(\text{MSE}(n)\sim cn^{-1}\) as \(n\) goes to infinity, where \(c\) is a constant. It also gives  an explicit
geometric bound on the Wasserstein distance between the distribution of \(\X_{n}\)
and \(N({\bff 0},\V)\), and  an explicit bound on the mean square error of a related estimator of \(m\).  Section~\ref{se:TheoExamples} gives examples and a numerical simulation, and shows that, under certain conditions, the total time needed by our method to achieve a given  standarized mean square error is \({O}^{*}(d^{2})\). Concluding remarks are given in a closing section.

An introduction to MCMC methods can be found in~\cite{dellaportas2003introduction}.
Our proof-techniques are based on coupling arguments. Conductance
techniques can also  be used to analyse mixing properties of Markov chains~(see
e.g. \cite{sinclair1992improved,kahale1997semidefinite,diaconis2009markov}).  Chernoff bounds for reversible discrete Markov chains in terms of the spectral gap have been established in~\cite{kahale1997large,gillman1998chernoff}. Previous theoretical results on  the performance of   hit-and-run algorithms have focused on their mixing properties (see~\cite{VempalaCousins2016,BRS1993hit} and references therein). For instance, after appropriate preprocessing, the hit-and-run algorithm for sampling from a convex body~\cite{lovasz1999hit} produces an approximately uniformly distributed
sample point  after \(O^{*}(d^{3})\) steps. For general log-concave functions,  after appropriate preprocessing~\cite{lovaszVempala2006},  the "hit-and-run" algorithm   mixes in  \(O^{*}(d^{4})\) steps. Note that, while the algorithms in~\cite{lovasz1999hit,lovaszVempala2006} require a pre-processing phase to make the target distribution "well-rounded", our method does not.   When  \(\V\) is positive definite,  the Metropolis and Gibbs algorithms,  and an  algorithm for sampling from general log-concave functions using a Langevin stochastic differential equation~\cite{Moulines2016sampling}, could be used to approximately sample from
  \(N(0,V)\), but these algorithms require the calculation of \(\V^{-1}\). Standard algorithms for inverting a matrix take \(\Theta(d^{3})\) time and  \(\Theta(d^{2})\) space, however, and so the pre-processing cost of these algorithms is as high as the Cholesky decomposition cost.   Omitted  proofs are in the appendix.   
   
\section{Motivation, notation and general properties}\label{se:Notation}
We motivate~\eqref{eq:xndefsimple} by assuming that \(\V\) is positive definite, which implies the existence of a lower-triangular matrix
  \(\A\)  satisfying~\eqref{eq:aat} and such that  \(\A\) and \(\V\) have the same  first column~\cite[Subsection 2.3.3]{glasserman2004Monte}. Let \(\Z\sim N(\bff0,\I)\), where \(\I\) is the \(d\times d\) identity matrix, and \(g\)  a standard Gaussian random variable independent of \(\Z\). Set \(\Z'=\Z+(g-\e_{1}^{T}\Z)\e_{1}\), and let  \(\X=\A\Z\) and \(\X'=\A\Z'=\X+(g-\e_{1}^{T}\Z)(\A\e_{1}\)).
Note that  \(\Z'\) is obtained from \(\Z\) by replacing its first component  \(\e_{1}^{T}\Z\)  with \(g\), and so     \(\Z'\sim N(\bff0,\I)\). Hence \(\X\sim \X'\sim N({\bff 0},\V)\). But, since \(\A\e_{1}\) (resp. \(\V\e_{1}\)) is the first column of \(\A\) (resp. \(\V\)), \(\A\e_{1}=\V\e_{1}\).  Furthermore,   \(\e_{1}^{T}\A=\e_{1}^{T}\)since
the first line of \(\A\) is \(\e_{1}^{T}\),  and so  \(\e_{1}^{T}\X=\e_{1}^{T}\Z\).  Thus
\begin{equation}\label{eq:xx'}
\X'=\X+(g-\e_{1}^{T}\X)(\V\e_{1}), 
\end{equation}and so the RHS of~\eqref{eq:xx'} is a centered Gaussian vector with covariance matrix \(\V\).   \eqref{eq:xndefsimple} is obtained from~\eqref{eq:xx'} by replacing \(\e_{1}\),
 \(g\), \(\X\) and \(\X'\) with \(\e_{i_{n}}\), \(g_{n}\), \(\X_{n}\) and \(\X_{n+1}\),
 respectively.

We now describe a generic standard hit-and-run algorithm  to approximately sample from a real-valued density function \(f\) on \(\mathbb{R}^{d}\). First,  choose \(\X^{\text{HR}}_{0}\) from a certain distribution. If we are currently at point \(\X^{\text{HR}}_{n}\), we first choose a random vector \({\bff u}_{n}\in\mathbb{R}^{d}\) according to a certain distribution, and then set \begin{equation*}
\X^{\text{HR}}_{n+1}=\X^{\text{HR}}_{n}+g^{\text{HR}}_{n}{\bff u}_{n}, 
\end{equation*}where  \(g_{n}^{\text{HR}}\)  is a random variable whose density at \(t\in\mathbb{R}\) is  proportional to \(f(\X^{\text{HR}}_{n}+t{\bff u}_{n})\). 

If \(\V\) is positive definite, the density \(f(\x)\) of \(N(\bff0,V)\) at \(\x\in\mathbb{R}^{d}\) is  \(\exp(-\x^{T}\V^{-1}\x/2)\), up to a multiplicative constant. In the standard hit and run algorithm, \({\bff u}_{n}\) is a uniformly distributed unit vector. However, if we set \(\X^{\text{HR}}_{0}=\bff0\) and \({\bff u}_{n}=\V\e_{i_{n}}\), it follows after some  calculations that \(g^{\text{HR}}_{n}\sim N(-\e_{i_{n}}^T\X^{\text{HR}}_{n},1)\). Thus, we can  choose \(g^{\text{HR}}_{n}=g_{n}-\e_{i_{n}}^T\X^{\text{HR}}_{n}\), which implies by induction that \(\X_{n}^{\text{HR}}=\X_{n}\).     
   
If \(\x\)  is a \(d\)-dimensional vector, denote by \(||\x||\)   the \(l_{2}\)-norm
of \(\x\).  For  any \(d\times d\) matrix \(\A\), the matrix  \(\A^{T}\V\A\) is positive semi-definite. Let \begin{displaymath}
||\A||=||\A||_{\V}=\sqrt{\tr(\A^{T}\V\A)}
\end{displaymath} be the Frobenius norm of the matrix   \(\sqrt{\V}\A\). If \(\A\) and \(\B\) are symmetric \(d\times d\) matrices, we say that \(\A\leq \B\) if \(\B-\A\) is positive semi-definite, and we denote by \(\lambda_{\max}(\A)\) the largest eigenvalue of \(\A\). If \(\Z\) is a   centered  \(d\)-dimensional random vector such that \(\E(||\Z||^{2})\) is finite, let  \(\cov(\Z)=\E(\Z\Z^{T})\) denote the covariance matrix of \(\Z\).    

For \(1\leq i\leq d\), let \(\f_{i}=\sqrt{\V}\e_{i}\) and  \(\Pf_{i}=\I-\f_{i}\f_{i}^{T}\).
 Note that \(||\f_{i}||^{2}=\e_{i}^{T}\V\e_{i}=1\). Thus \(\f_{i}\) is a unit vector   and  \(\Pf_{i}\) is a projection matrix, i.e. \(\Pf_{i}^{2}=\Pf_{i}\). 
Define the random sequence of \(d\)-dimensional vectors \(\Y_{n}\), \(n\geq0\),
as follows:
\(\Y_{0}=\bff0\) and \begin{equation*}
 \Y_{n+1}= \Pf_{i_{n}}\Y_{n}+g_{n}\f_{i_{n}}.
 \end{equation*}
By rewriting~\eqref{eq:xndefsimple} as   \begin{equation*}\label{eq:xndefff}
\X_{n+1}=(\I-\V\e_{i_{n}}\e_{i_{n}}^{T})\X_{n}+g_{n}(\V\e_{i_{n}}),
\end{equation*}  it can be shown by induction that \(\X_{n}=\sqrt{\V}\Y_{n}\).

For \(0\leq m\leq  n\), let  \(\M_{m,n}=\Pf_{i_{n-1}}\Pf_{i_{n-2}}\cdots \Pf_{i_{m}}\), with \(\M_{n,n}=\I\), and let \(\M_{n}=\M_{0,n}\).  Let \(\Z_{0}\) be a  \(d\)-dimensional vector of  independent standard Gaussian random variables which is independent of  the sequence \((g_{n},i_{n})\), \(n\geq0\). For \(n\geq1\),
let \begin{equation}
\label{eq:wWn+1}
 \Z_{n}= \Y_{n}+\M_{n}\Z_{0}.
 \end{equation}
Since \(\lambda_{\max}(\A)\leq\tr(\A)\) for a positive semi-definite matrix \(\A\), the following lemma implies that, if the sequence  \((i_{k})\), \(k\ge 0\), is  deterministic, then \(\)\(\X_{n}\) is centered Gaussian and   \begin{equation*}
 \lambda_{\max}(\V-\cov(\X_{n}))\le||\M_{n}||^{2}. 
\end{equation*} As a consequence, any entry of \(\V-\cov(\X_{n})\) is upper-bounded, in absolute value, by \(||\M_{n}||^{2}\).   
 \begin{lemma}
\label{le:deterministic}
If the sequence  \((i_{k})\), \(0\le k\leq n-1\), is  deterministic, then, for \(0\le m\le n\), \(\X_{n}\) and \(\Y_{n}\) are centered Gaussian vectors,    \(\Z_{n}\sim N(\bff0,\I)\), and \begin{equation}\label{eq:covZ}
\E(\Z_{n}\Z^{T}_{m})=\M_{m,n}.
\end{equation}
Furthermore,   \begin{equation}\label{eq:rn}
\cov(\X_{n})=\V-\sqrt{\V}\M_{n}\M_{n}^{T}\sqrt{\V}, 
\end{equation} \(\cov(\X_{n})\leq \V\), and \begin{equation}\label{eq:trRn}
\tr(\V-\cov(\X_{n}))=\E(||\X_{n}-\sqrt{\V}\Z_{n}||^{2})=||\M_{n}||^{2}.
\end{equation}
\end{lemma}
   Lemma~\ref{le:deterministic} forms the basis for
 the proofs of our main results. Indeed, if \(\M_{m,n}\) goes to \(\bff0\) as  \(n-m\)
 goes to infinity then, by~\eqref{eq:rn},  \(\cov(\X_n)\)  converges to \(\V\)
 as \(n\) goes to infinity. Furthermore, if both \(m\) and \(n-m\) are sufficiently
 large then, by \eqref{eq:covZ}, \(\Z_{n}\)  and \(\Z_{m}\) are nearly independent and, by  \eqref{eq:wWn+1}, \(\Y_{m}\) (resp. \(\Y_{n}\)) is close
 to  \(\Z_{m}\) (resp. \(\Z_{n}\)). Thus,  \(\Y_{m}\)  and \(\Y_{n}\) are nearly independent, as well, and so are \(\X_{m}\) and \(\X_{n}\). These arguments are informal
since we have not defined the terms "nearly independent'' and ''close'',
but give  intuition behind the proofs of Theorems~\ref{th:trSum}
and \ref{th:MSE}.

 Lemma~\ref{le:Wasserstein} below generalizes some   results of Lemma~\ref{le:deterministic}
 when the sequence  \((i_{k})\), \(k\geq0\), is    random. 
\begin{lemma}\label{le:Wasserstein} If the sequence  \((i_{k})\), \(k\geq0\), is  deterministic or random, the
 quadratic Wasserstein distance between the distribution of \(\X_{n}\) and  \(N({\bff 0},\V)\)  is at most \(\sqrt{\E(||\M_{n}||^{2})}\).
 Furthermore,     \(\Z_{n}\sim N(\bff0,\I)\),  \(\X_{n}\) is centered, \(\cov(\X_{n})\leq \V\), and  \begin{equation}\label{eq:trRnRandom}
\tr(\V-\cov(\X_{n}))=\E(||\M_{n}||^{2}).
\end{equation}  \end{lemma}
\commentt{\begin{proof}}{\proof{Proof.}}
  By Lemma~\ref{le:deterministic}, conditioning on \(i_{0},\dots,i_{n}\),  \(\Z_{n}\sim N({\bff0},\I)\). Thus, the unconditional distribution of   \(\Z_{n}\) is \(N({\bff0},\I)\), and   \(\sqrt{{V}}\Z_{n}\sim N({\bff0},\V)\). On the other hand, by~\eqref{eq:trRn},\begin{displaymath}
\E(||\X_{n}-\sqrt{\V}\Z_{n}||^{2}|i_{0},\dots,i_{n})
=||\M_{n}||^{2},
\end{displaymath}and so, by the tower law, \begin{equation*}
\E(||\X_{n}-\sqrt{\V}\Z_{n}||^{2})= \E(||\M_{n}||^{2}).
\end{equation*}By~\eqref{eq:defWasserstein}, it follows that the quadratic Wasserstein distance between the distribution of \(\X_{n}\) and \(N(\bff0,V)\) is at most \(\sqrt{ \E(||\M_{n}||^{2})}\). 
On the other hand, it follows from Lemma~\ref{le:deterministic} that 
\(\E(\X_{n}|i_{0},\dots,i_{n})=\bff0\). By the tower law, we infer that \(\X_{n}\) is centered. Similarly, by Lemma~\ref{le:deterministic},\begin{displaymath}
\E(\X_{n}\X_{n}^{T}|i_{0},\dots,i_{n})\le \V.
\end{displaymath}Hence, by the tower law,  \(\E(\X_{n}\X_{n}^{T})\le \V\), and so   \(\cov(\X_{n})\leq \V\). Once again,~\eqref{eq:trRnRandom} follows from~\eqref{eq:trRn} by the tower law.
\commentt{\end{proof}}{\Halmos\endproof} 
\section{Upper bound on the Wasserstein distance} \label{se:random}We first
 show the following lemma. 
\comment{\begin{lemma}\label{le:fundamental}
For 
 \(n\geq1\), 
\begin{equation}\label{eq:MnfiRnd}
||\sqrt{\V}\E(\v\vv_{n})||\leq (d/n)^{1/2},
\end{equation}and\begin{equation}
 \label{eq:fiTMnvRnd}
\sum^{\infty}_{n=0}||\sqrt{\V}\E(\vv_{n})||^{2}\leq d.
\end{equation}
\comment{
For \(j\geq0\),
\begin{equation}\label{eq:Mnfi}
||\M^{n}(\M^T)^{j}\f_{i}||\leq (d/n)^{1/2}.
\end{equation}
}
\end{lemma}
\commentt{\begin{proof}}{\proof{Proof.}}
A simple calculation shows that \(\E(\e_{i_{n}}\e_{i_{n}}^{T})=d^{-1}\I\), and so \(\E(\Pf_{i_{n}})=\I-d^{-1}\V\).
 Hence  \begin{equation}
\E(\vv_{n+1})=(\I-d^{-1}\V)\E(\vv_{n}).
\end{equation} Assume now that \(\vv\) is an eigenvector of  \(\V\) with eigenvalue \(\lambda_{i}\).
Since \(\sqrt{\V}\vv=\sqrt{\lambda_{i}}\,\vv\),
 \begin{equation}
\sqrt{\V}\E(\vv_{n})=\sqrt{\lambda_{i}}(1-\frac{\lambda_{i}}d)^{n}\,\vv,
\end{equation}
and 
\begin{equation}
||\sqrt{\V}\E(\vv_{n})||^{2}=\lambda_{i}(1-\frac{\lambda_{i}}d)^{2n}.
\end{equation}
Since \(\x(1-\x)^{n}\leq1/(n+1)\) for \(0\leq \x\leq1\), it follows from linearity
of expectations that  \ref{eq:MnfiRnd} and \ref{eq:fiTMnvRnd} hold for any unit-vector \(\vv\).
\commentt{\end{proof}}{\Halmos\endproof}
}

 \begin{lemma}\label{lemma:trProjection}
If \(\Pf\) is a  \(d\times d\) projection matrix and \(\A\) is a \(d\times d\) matrix,
then \(||\A\Pf||\leq||\A||\).   
\end{lemma}
\commentt{\begin{proof}}{\proof{Proof.}}
 Let \({\bff H}=\A^{T}\V\A\). Since \(\tr(\bff BC)=\tr(CB)\), \( \tr(\bff PHP)=\tr(HP)=\tr(PH)\), and
so   \(\tr(\bff H)-\tr(PHP)=\tr((\I-\Pf)H(\I-\Pf))\).  Since \(\bff H\) is positive
semi-definite, so is  \((\I-\Pf)\bff H(\I-\Pf)\), and so  \(\tr(\bff PHP)\le\tr(H)\). Equivalently,
 \(||\A\Pf||^{2}\leq||\A||^{2}\).
This concludes the proof.
\commentt{\end{proof}}{\Halmos\endproof}
Under the conditions stated in Theorem~\ref{th:trSum} below, by an argument similar to that surrounding Lemma~\ref{le:deterministic}, it follows from~\eqref{eq:trRnRandomUnif}
that each entry of
the matrix \(\V-\cov(\X_{n})\) is at most \(d^{2}/n\) in absolute value.
\begin{theorem}\label{th:trSum} Assume     that \(i_{n}\), \(n\geq0\),  are independent random variables   uniformly distributed over \(\{1,\dots,d\}\).
For \(n\geq1\), the
 quadratic Wasserstein distance between the distribution of \(\X_{n}\) and  \(N({\bff 0},\V)\)  is at most \(d/\sqrt{n}\),  
\begin{equation}\label{eq:sumtr}
\sum^{n}_{j=0}\E(||\M_{j}||^{2})\leq d^{2},
\end{equation}
 and the sequence \(\E(||\M_{j}||^{2})\) is decreasing. Furthermore, for \(n\ge1\),    \(\X_{n}\) is centered, \(\cov(\X_{n})\leq \V\), and 
 \begin{equation}
\label{eq:trRnRandomUnif}
\tr(\V-\cov(\X_{n}))\le \frac{d^{2}}{n}.
\end{equation} \end{theorem}
 \commentt{\begin{proof}}{\proof{Proof.}}For any non-negative integer \(j\), \begin{eqnarray*}
\E(\e_{i_{j}}\e_{i_{j}}^{T})&=&d^{-1}\sum^{d}_{i=1}\e_{i}\e_{i}^{T}\\&=&d^{-1}\I.
\end{eqnarray*}Thus,
 \begin{eqnarray*}
\E(\f_{i_{j}}\f_{i_{j}}^{T})&=&\sqrt{\V}\E(\e_{i_{j}}\e_{i_{j}}^{T})\sqrt{\V}\\&=&d^{-1}\V,
\end{eqnarray*}and so 
\begin{equation*}\label{eq:expectedPi}
\E(\Pf_{i_{j}})=\I-d^{-1}\V.
\end{equation*} 
Let \(\vv\) be a unit \(d\)-dimensional vector. For \(j\geq0\), set  \(\vv_{j}=\M_{j}\vv\).  Since \(\Pf_{i}\) is a projection and \(\vv_{j+1}=\Pf_{i_{j}}\vv_{j}\)
for
\(j\geq0\), it follows that
\begin{eqnarray*}
\E(||\vv_{j+1}||^{2})&=&\E(\vv_{j}^{T}\Pf_{i_{j}}\vv_{j})\\
&=&\E(||\vv_{j}||^{2})-d^{-1}\E(\vv_{j}^{T}\V\vv_{j}).
\end{eqnarray*}
Hence
\begin{equation}\label{eq:dvjTVvj}
d^{-1}\E(\vv_{j}^{T}\V\vv_{j})=\E(||\vv_{j}||^{2})-\E(||\vv_{j+1}||^{2}).
\end{equation}As \(\vv_{0}=\vv\), we conclude that
\begin{equation*}
\sum^{n}_{j=0}\E(\vv_{j}^{T}\V\vv_{j})\leq d.
\end{equation*}But\begin{equation}\label{eq:vjTVvj}
\vv_{j}^{T}\V\vv_{j}=\vv^{T}\M_{j}^{T}\V\M_{j}\vv,
\end{equation}
and so, for any unit vector \(\vv\),
 \begin{equation*}
\vv^{T}(\E(\sum^{n}_{j=0}\M_{j}^{T}\V\M_{j}))\vv\leq d.
\end{equation*} Thus, any diagonal entry of  the  matrix \(\E(\sum^{n}_{j=0}\M_{j}^{T}\V\M_{j})\) is at most \(d\). Hence, \begin{equation*}
\tr(\E(\sum^{n}_{j=0}\M_{j}^{T}\V\M_{j}))\leq d^{2},
\end{equation*}which implies~\eqref{eq:sumtr}. On the other hand, since \(\M_{j+1}^{T}=\M_{j}^{T}\Pf_{i_{j}}\), Lemma~\ref{lemma:trProjection} shows that 
\(||\M_{j+1}^{T}||\le||\M_{j}^{T}||
\). Hence the sequence \(\E(||\M_{j}^{T}||^{2})\) is  decreasing.  Since \(\M_{j}\) and \(\M^{T}_{j}\) have the same distribution, \(\E(||\M_{j}^{T}||^{2})=\E(||\M_{j}||^{2})\). Thus,
 the sequence \(\E(||\M_{j}||^{2})\) is decreasing as well and so, by~\eqref{eq:sumtr},   \(nE(||\M_{n}||^{2})\leq d^{2}\). We conclude the proof using Lemma~\ref{le:Wasserstein}.
  \commentt{\end{proof}}{\Halmos\endproof} \comment{
Let \begin{equation}
||\A||=\tr(\sqrt{\A^{T}\A})
\end{equation}
be the trace norm of \(\A\).
\begin{lemma}
\begin{equation}
2||\A\B||\leq\tr(\A^{T}\A)+\tr(\B^{T}\B).
\end{equation}
\end{lemma}
 \commentt{\begin{proof}}{\proof{Proof.}}
Let \(\V'=\sqrt{\A^{T}\V\A}\), where \(\V=\B^{T}\B\). Then \begin{eqnarray*}
2\x^{T}\V'\x&=&2\x^{T}\V'\A^{-1}Ax\\
&\le&||Ax||^{2}+||{\A^{-1}}^{T}\V'\x||^{2}
\end{eqnarray*}
and so \begin{equation}
2\V'\leq \A^{T}\A+\V'{\A^{-1}}{\A^{-1}}^{T}\V' 
\end{equation}
Hence \begin{equation}
2\tr(\V')\leq\tr(\A^{T}\A)+\tr(\V'{\A^{-1}}{\A^{-1}}^{T}\V')
\end{equation}
Since \begin{equation}
\tr(\V'{\A^{-1}}{\A^{-1}}^{T}\V')=\tr(\V)
\end{equation}
\commentt{\end{proof}}{\Halmos\endproof}
}
\comment{
\begin{lemma}
\begin{equation}
\sum^{\infty}_{j=0}\E(||\E(\X_{n+j}\X^{T}_{j}|(i))||)\leq4 d^{2}.
\end{equation}
\end{lemma}
 \commentt{\begin{proof}}{\proof{Proof.}}
 By Eq.~\ref{eq:Wn+1}, 
 \begin{equation*}
\E(\Y_{n+j+1}|\Y_{n},(i))=\Pf_{i_{n+j}}\E(\Y_{j+n}|\Y_{n},(i))
\end{equation*} 
and so \begin{equation*}
\E(\Y_{n+j}|\Y_{n},(i))=\M_{n,n+j}\Y_{n}.
\end{equation*}
Thus, by Eq.~\ref{eq:r'nd},
\begin{eqnarray*}
\E(\Y_{n+j}\Y^{T}_{n}|(i))&=&\E(\M_{n,n+j}\Y_{n}\Y_{n}^{T}|(i))\\
&=&\M_{n,n+j}(\I-\M_{n}\M_{n}^T)\\
&=&\M_{n,n+j}-\M_{n+j}\M_{n}^T.
\end{eqnarray*}
Hence,
\begin{equation}
\E(\X_{n+j}\X^{T}_{n}|(i))=\sqrt{\V}\M_{n,n+j}\sqrt{\V}-\sqrt{\V}\M_{n+j}\M_{n}^T\sqrt{\V}.
\end{equation}
Let \(j'=\lfloor j/2\rfloor\), \(\A=\sqrt{\V}\M_{n+j',n+j}\) and \(\B=\M_{n,n+j'}\sqrt{\V}\).
Thus,
\begin{equation}
2||\sqrt{\V}\M_{n,n+j}\sqrt{\V}||\leq\tr(\M_{n+j',n+j}^{T}\V\M_{n+j',n+j}+\M_{n,n+j'}^{T}\V\M_{n,n+j'})
\end{equation}
and so
\begin{equation}
||\sqrt{\V}\E(\M_{n,n+j})\sqrt{\V}||\leq \E(\tr(R_{ j'})).
\end{equation} 
Similarly,
\begin{equation}
||\sqrt{\V}\E(\M_{n+j}\M_{n}^T)\sqrt{\V}||\leq \E(\tr(R_{ j'})).
\end{equation}
\commentt{\end{proof}}{\Halmos\endproof}
}

\comment{
\begin{lemma}
Let \((\X, \X')\) be a Gaussian vector with \(\X\sim \X'  \sim N({\bff 0},\V)\)  and \(\E(UU'^{T})=\Gamma\). Assume that \(\sqrt{\Gamma^{T}\Gamma}\leq \V\). Let  \(f\) be a function such that  \(\E((f(\Y)-f(\Y'))^{2})\leq \kappa \E(||\Y-\Y'||^{2})\)
if  \((\Y, \Y')\) is a  Gaussian vector with     \(\cov( \Y)\leq \V\) and      \(\cov( \Y')\leq \V\). Then   \begin{equation}\label{eq:UpperBoundCovGaussianBis}
\cov(f(\X),f(\X'))\leq \kappa\,\tr(\sqrt{\Gamma^{T}\Gamma}).
\end{equation}   
\end{lemma}
 \commentt{\begin{proof}}{\proof{Proof.}}Let \(\Gamma=PV'\) be a polar decomposition of \(\Gamma\), where  \(\Pf\) is an orthogonal matrix and \(\V'=\sqrt{\Gamma^{T}\Gamma}\)  . Let \({\bff G}_{0}\),  \({\bff G}_{1}\) and \({\bff G'}_{1}\) be independent Gaussian random variables such that  \({\bff G}_{0}\sim N({\bff 0},\V')\),   \({\bff G}_{1}\sim N({\bff 0},\V-\Gamma \Pf^{T})\), and    \( {\bff G'}_{1}\sim N({\bff 0},\V-\V')\). Let \(\Y=P{\bff G}_{0}+{\bff G}_{1}\) and \(\Y'={\bff G}_{0} +{\bff G'}_{1}\). Then     \(\Y\sim \Y'\sim N({\bff 0},\V)\), and\begin{displaymath}
\E(WW'^{T})=\Gamma.
\end{displaymath} Thus \begin{eqnarray*}\cov(f(\X),f(\X'))
&=&\cov(f(\Y),f(\Y'))\\
&=&\cov(f(\Y)-f({\bff G}_{1}),f(\Y')-f({\bff G'}_{1}))\\
&\le&\frac{1}{2}\E((f(\Y)-f({\bff G}_{1}))^{2}+(f(\Y')-f({\bff G'}_{1}))^{2})\\
&\le&\frac{\kappa}{2}\E(||P{\bff G}_{0}||^{2}+||{\bff G}_{0}||^{2})\\
&=&\frac{\kappa}{2}(\tr(PV'\Pf^{T})+\tr(\V'))\\
&=&\kappa \tr(\V').
\end{eqnarray*}       
\commentt{\end{proof}}{\Halmos\endproof}
}

\comment{
On the other hand, since \begin{equation}
||\sqrt{\V}\vv_{n}||^{2}=\sum^{d}_{i=1}(\f_{i}^{T}\vv_{n})^{2},
\end{equation}
\begin{equation}
\sum^{\infty}_{n=0}\sum^{d}_{i=1}\E((\f_{i}^{T}\vv_{n})^{2})\leq d.
\end{equation}
and so
\begin{equation}
\sum^{\infty}_{n=0}\sum^{d}_{i=1}\E((\f_{i}^{T}\M_{n}\vv)^{2})\leq d.
\end{equation}
 Thus,
\begin{equation}
\sum^{\infty}_{n=0}\sum^{d}_{i=1}\E(||\M_{n}^{T}\f_{i}||^{2}))\leq d^{2}.
\end{equation}
Equivalently, 
\begin{equation}
\sum^{\infty}_{n=0}\tr(\sqrt{\V}\M_{n}\M_{n}^{T}\sqrt{\V})\leq d^{2}.
\end{equation}
\commentt{\end{proof}}{\Halmos\endproof}
}

\comment{
\begin{lemma}For \(1\leq k,l\leq d\)
\begin{equation}
\sum^{\infty}_{j=0}||\E(\X_{n+j}\X^{T}_{j})||_{1}\leq4 d^{3}.
\end{equation}
\end{lemma}
 \commentt{\begin{proof}}{\proof{Proof.}}
 By Eq.~\ref{eq:Wn+1}, 
 \begin{equation*}
\E(\Y_{n+j+1}|\Y_{n},(i))=\Pf_{i_{n+j}}\E(\Y_{j+n}|\Y_{n},(i))
\end{equation*} 
and so \begin{equation*}
\E(\Y_{n+j}|\Y_{n},(i))=\M_{n,n+j}\Y_{n}.
\end{equation*}
Thus, by Eq.~\ref{eq:r'nd},
\begin{eqnarray*}
\E(\Y_{n+j}\Y^{T}_{n},(i))&=&\E(\M_{n,n+j}\Y_{n}\Y_{n}^{T},(i))\\
&=&\M_{n,n+j}(\I-\M_{n}\M_{n}^T).
\end{eqnarray*}
Hence,
\begin{equation*}
\E(\X_{n+j}\X^{T}_{n},(i))=\sqrt{\V}\M_{n,n+j}(\I-\M_{n}\M_{n}^T)\sqrt{\V}
\end{equation*}
and so
\begin{equation}\label{eq:ekTE}
\e_{k}^{T}\E(\X_{n+j}\X^{T}_{n},(i))\e_{l}=\f_{k}^{T}\M_{n,n+j}\f_{l}-\f_{k}^{T}\M_{n+j}\M_{n}^Tf_{l}.
\end{equation}
On the other hand, by the Cauchy-Schwartz inequality, \begin{eqnarray*}
|\f_{k}^{T}\M_{n,n+j}\f_{l}|&\leq&||\M_{n+\lfloor j/2\rfloor,n+j}^{T}\f_{k}||\,||\M_{n,n+\lfloor j/2\rfloor}\f_{l}||\\&\leq&\frac{1}{2}(||\M_{n+\lfloor j/2\rfloor,n+j}^{T}\f_{k}||^{2}+||\M_{n,n+\lfloor j/2\rfloor}\f_{l}||^{2}).
\end{eqnarray*}
Thus,\begin{equation}
\E(|\f_{k}^{T}\M_{n,n+j}\f_{l}|)\leq \frac{1}{2}\E(||\M_{\lfloor j/2\rfloor}\f_{k}||^{2}+||\M_{\lfloor j/2\rfloor}\f_{l}||^{2})
\end{equation}
and so 
\begin{equation}
\E(||\M_{n,n+j}||_{1})\leq dE(\tr(R_{\lfloor j/2\rfloor})).
\end{equation}
\commentt{\end{proof}}{\Halmos\endproof}

Let \(f\) be a real-valued function on \(\mathbb{R}^{d}\). Assume that, for
any \(d\times d\) matrices \(\V\), \(\V'\) and \(\Psi\), the functions \(g(\V)=\E(f(\X))\) and
\(h(\V,\V',\Psi)=\cov(f(\X),f(\X'))\) if \((\X,\X')\sim N({\bff 0},\begin{pmatrix}\V & \Psi \\
\Psi^{T} & \V' \\
\end{pmatrix})\) exist.  Assume further that \(|g(\V)-g(\V')|\leq \kappa\, \tr(\V-\V')\)
if \(\V-\V'\) is positive semi-definite.
}
\comment{
\begin{lemma}\label{le:varDecomp}
Let \({\bff G}_{0}\) and \({\bff G}_{1}\) be independent \(d\)-dimensional random   variables and \(\Y\) a real-valued random variable with finite second moment.  Then \begin{equation}\label{eq:varDecomp}
\var(\E(\Y|{\bff G}_{0}))+\var(\E(\Y|{\bff G}_{1}))\le\var(\Y).
\end{equation}
\end{lemma}
 \commentt{\begin{proof}}{\proof{Proof.}}
Let  \(\Z=\Y+\E(\Y)-\E(\Y|{\bff G}_{0})-\E(\Y|{\bff G}_{1})\). By the tower rule, \(\E(\Z)=0\). On the other
hand, since \({\bff G}_{0}\) and \({\bff G}_{1}\) are independent, \begin{equation*}
\E(\E(\Y|{\bff G}_{1})|{\bff G}_{0})=\E(\Y).
\end{equation*}
Thus \(\E(\Z|{\bff G}_{0})=0\), and so \(\E(ZE(\Y|{\bff G}_{0}))=0\). Hence \(
\cov(\Z,\E(\Y|{\bff G}_{0}))=0
\). Similarly,  \(
\cov(\Z,\E(\Y|{\bff G}_{1}))=0
\). Hence the four 
components of the following decomposition of \(\Y\)\begin{displaymath}
 \Y=\Z-\E(\Y)+\E(\Y|{\bff G}_{0})+\E(\Y|{\bff G}_{1}) 
\end{displaymath}
have a null pair-wise covariance, and so \begin{displaymath}
\var(\Y)=\var(\Z)+\var(\E(\Y|{\bff G}_{0}))+\var(\E(\Y|{\bff G}_{1})).
\end{displaymath}
Hence~\eqref{eq:varDecomp}.\commentt{\end{proof}}{\Halmos\endproof}
}
\comment{
\begin{theorem}
Let \(f\) be a Borel-measurable function on \(\mathbb{R}\) such that  \(\E(|f(g)|g^{2})\)
is finite and \(|f(\x)-f(0)|\leq \kappa|\x|\) for \(|\x|\leq\epsilon\), where    \(\kappa\) and \(\epsilon\) are positive constants and \(g\) is a standard Gaussian random
variable. Then \(\sigma \mapsto \E(f(\sigma g))\) is Lipschitz for \(\sigma\in[-1,1]\).
\end{theorem}
 \commentt{\begin{proof}}{\proof{Proof.}}
Assume without loss of generality that \(f(0)=0\) and \(\epsilon\leq1\). Since\begin{equation}
|f(\x)|N'( \frac{\x}{\sigma})\leq (\kappa\epsilon+(\frac{\x}{\epsilon})^{2}|f(\x)|)N'( \x),
\end{equation}
\(\E(f(\sigma g))\) exists for \(\sigma\in[-1,1]\). On the other hand,\begin{equation}
\frac{\partial}{\partial \sigma}(\frac{1}{\sigma}N'( \frac{\x}{\sigma}))=(\frac{\x^{2}}{\sigma^{4}}-\frac{1}{\sigma^{2}})N'( \frac{\x}{\sigma}).
\end{equation}
Since  \(y^{4}N'(y)\) and  \(y^{2}N'(y)\) are upper  bounded
by \(16N'(1)\), it follows that, for \(\sigma \leq\epsilon\leq \x\), \begin{equation}|
\frac{\partial}{\partial \sigma}(\frac{1}{\sigma}N'( \frac{\x}{\sigma}))|\leq16(\frac{\x^{2}}{\epsilon^{4}}+\frac{1}{\epsilon^{2}})N'( \frac{\x}{\epsilon})
\end{equation}
and so,  for \(\x\geq\epsilon \), and \( 0<\sigma\leq 1\), \begin{equation}|
\frac{\partial}{\partial \sigma}(\frac{1}{\sigma}N'( \frac{\x}{\sigma}))|\leq32\frac{\x^{2}}{\epsilon^{4}}N'(\x).
\end{equation}  We conclude that \(|f(\x)
\frac{\partial}{\partial \sigma}(\frac{1}{\sigma}N'( \frac{\x}{\sigma}))|\)
is uniformly integrable over \(\mathbb{R}\).
\commentt{\end{proof}}{\Halmos\endproof}
}
\section{Bounding  the mean square error}\label{se:Lipschitz}
We now define the class of  \((\kappa,\gamma,\bff W)\)-Lipschitz  functions, with \(\kappa >0\) and \(\gamma\in(0,1]\). 
\begin{definition}
Let \(\bff W\) be a \(d\times d\) positive semi-definite matrix. We say that a real-valued Borel function   \(h\) of \(d\) variables is \((\kappa,\gamma,\bff W)\)-Lipschitz if\begin{equation}\label{eq:defkappaLips}
\E((h(\X)-h(\X'))^{2})\leq \kappa^2( \E(||\X-\X'||^{2}))^{\gamma}
\end{equation}
for any centered Gaussian column vector \(\begin{pmatrix}\X \\
\X' \\
\end{pmatrix}\)   with     \(\cov(\X)\le \bff W\) and
  \(\cov(\X')\le\bff W\), where \(\X\) and \(\X'\) are \(d\)-dimensional. \end{definition}
We say that  a function \(h\) is   \((\kappa,\bff W)\)-Lipschitz if it is    \((\kappa,1,\bff W)\)-Lipschitz. For instance, if  \(h\)  a real-valued \(\kappa\)-Lipschitz function on
\(\mathbb{R}^{d}\), i.e. \(|h(\x)-h(\x')|\leq\kappa||\x-\x'||\) for \(\x\), \(\x'\)
in \(\mathbb{R}^{d}\), then \(h\) is \((\kappa,\bff W)\)-Lipschitz for
any  \(d\times d\) positive semi-definite matrix \(\bff W\). The following lemma gives an example of a \((\kappa,\bff W)\)-Lipschitz function in \(\mathbb{R}\) which is not  \(\kappa'\)-Lipschitz for any \(\kappa'>0\).      
\begin{lemma}\label{pr:GaussianLip1} Let  \(f(z)=e^{z}\). Then \(f\) is \((e^{\nu}\sqrt{4\nu +1},\nu)\)-Lipschitz
for \(\nu\geq0\).  
\end{lemma}

Let \(h\) be a  \((\kappa,\gamma,\V)\)-Lipschitz function.  Set \(m=\E(h(\X))\) and \(\Sigma^{2}=\var(h(\X))\), where \(\X\sim N({\bff 0},\V)\),  and denote by  \(\hat h\) the real-valued function on \(\mathbb{R}^{d}\) defined by \(\hat h(\x)=h(\sqrt{\V}\x)-m\).  Note that  \(\E(\hat h(\Z))=0\)  if \(\Z\sim N({\bff0},\I)\), since  \(\sqrt{\V} \Z\sim N({\bff 0},\V)\).   In particular, by Lemma~\ref{le:Wasserstein},   \(\E(\hat h(\Z_{j}))=0\) for \(j\geq0\).  
In other words, \(\E(h(\sqrt{V}\Z_{j}))=m\), and so the average of \(h(\sqrt{V}\Z_{j})\),
\(b\leq j\leq n-1\), where \(b\) is a burn-in period, is an unbiased estimator of \(m\). The variance of this estimator  equals \((n-b)^{-2}\E((\sum^{n-1}_{j=b}\hat h(\Z_{j}))^{2})\), which we
bound using  Lemma~\ref{le:barhZj} below. Choices for the parameters \(b\) and \(\delta\) will be given in the sequel.      
\begin{lemma}\label{le:barhZj}
Let   \(b\), \(n\) and \(\delta \) be  integers, with \(0\leq \delta \leq b<n\). Then   
\begin{equation*}
\E((\sum^{n-1}_{j=b}\hat h(\Z_{j}))^{2})\leq 4(n-b)\delta \Sigma^{2}+4\kappa^2\sum^{}_{b\leq j,\, j+\delta\leq l\leq n-1\ } \E(||\M_{j,l}||^{2\gamma}+||\M_{j,l}^{T}||^{2\gamma}).
\end{equation*}
\end{lemma}
For \(j\geq 0\), let\begin{equation}\label{eq:betajDef}
\beta_{j}=\hat h(\Z_{j})-\hat h(\Y_{j}),\text{ and }\beta=\sum^{n-1}_{j=b} \beta _{j}.
\end{equation}The second moments of \(\beta_{j}\) and of \(\beta\) can be bounded as follows.  
 \begin{lemma}\label{le:ebarh2}
 Let   \(b\) and \(n\) be  integers, with \(0\leq  b<n\).  Then \begin{equation*}\E({\beta_{j}}^{2})\leq\kappa^2E(||\M_{j}||^{2\gamma}),\end{equation*}and \begin{equation*}\label{eq:barh2random}
\E(\beta^{2})\leq (n-b)\kappa^2  \sum^{n-1}_{j=b}\E(||\M_{j}||^{2\gamma}).
\end{equation*}
 \end{lemma}
 \commentt{\begin{proof}}{\proof{Proof.}}
 Assume first that the sequence \(i_{0},\dots,i_{n-1}\), is deterministic.\
 For \(0\leq j\leq n-1\), 
\begin{eqnarray*}\label{eq:betaj}\E({\beta_{j}}^{2})
&=& \E(||h(\sqrt{\V}\Z_{j})-h(\X_{j})||^{2})\nonumber\\
&\leq& \kappa^2(\E(||\sqrt{\V}\Z_{j}-\X_{j}||^{2}))^{\gamma}\nonumber\\
&=& \kappa^2||\M_{j}||^{2\gamma}.
  \end{eqnarray*} The second equation follows from the relations \(\cov(\sqrt{\V}\Z_{j})=\V\) and \(\cov(\X_{j})\le\V\), and the last equation follows from~\eqref{eq:trRn}.   Thus, for any random sequence \(i_{0},\dots,i_{n-1}\), \begin{equation*}\E(\beta_{j}^{2}|i_{0},\dots,i_{j})\leq\kappa^2||\M_{j}||^{2\gamma}.\end{equation*}
The first inequality in the lemma then follows by taking expectations and using the tower law. The second inequality follows from the first one and the Cauchy-Schwartz inequality. 
\commentt{\end{proof}}{\Halmos\endproof}

Combining Lemmas~\ref{le:barhZj} and~\ref{le:ebarh2} yields the following.\begin{lemma}\label{le:MSEGen}Let  \(b\), \(n\) and \(\delta \) be  integers, with \(0\leq \delta \leq b<n\).  
If  \(i_{0},\dots,i_{n-1}\)  are independent random variables   uniformly distributed over \(\{1,\dots,d\}\), then \begin{equation*}
\E((\frac{\sum ^{n-1}_{j=b}h(\X_{j})}{n-b}-m)^{2})\leq\frac{1}{n-b}(8\delta \Sigma^{2}+18\kappa^2 \sum^{n-1}_{j=\delta }\E(||\M_{j}||^{2\gamma}).
\end{equation*}
\end{lemma}
\commentt{\begin{proof}}{\proof{Proof.}}
Since \(\M_{j,l}\sim \M_{l-j} \sim \M_{l-j}^{T} \),  for any fixed \(j\geq0\), \begin{displaymath}
 \sum_{l= j+\delta }^{n-1}\E(||\M_{j,l}||^{2\gamma})\leq \sum^{n-1}_{j=\delta }\E(||\M_{j}||^{2\gamma}),
\end{displaymath}
and \begin{displaymath}
 \sum_{l= j+\delta }^{n-1}\E(||\M_{j,l}^{T}||^{2\gamma})\leq \sum^{n-1}_{j=\delta }\E(||\M_{j}||^{2\gamma}).
\end{displaymath}Hence, by   Lemma~\ref{le:barhZj},  
\begin{equation*}
\E((\sum^{n-1}_{j=b}\hat h(\Z_{j}))^{2})\leq 4(n-b)\delta \Sigma^{2}+8(n-b)\kappa^2 \sum^{n-1}_{j=\delta }\E(||\M_{j}||^{2\gamma}).
\end{equation*}As \begin{displaymath}
(\sum^{n-1}_{j=b}\hat h(\Y_{j}))^{2}\leq2\beta^{2}+2(\sum^{n-1}_{j=b}\hat h(Z_{j}))^{2},
\end{displaymath} it follows by  Lemma~\ref{le:ebarh2} that\begin{equation*}
\E((\sum^{n-1}_{j=b}\hat h(\Y_{j}))^{2})\leq 8(n-b)\delta \Sigma^{2}+18(n-b)\kappa^2 \sum^{n-1}_{j=\delta }\E(||\M_{j}||^{2\gamma}).
\end{equation*}
Since \(\hat h(\Y_{j})=h(\X_{j})-m\), this concludes the proof.
 \commentt{\end{proof}}{\Halmos\endproof}
By applying Lemma~\ref{le:MSEGen} with \(b=\delta =0\) and using~\eqref{eq:sumtr}, we get the following upper bound on \(\text{MSE}(n)\). \begin{theorem}\label{th:MSE}Let \(h\) be a \((\kappa,\V)\)-Lipschitz function on \(\mathbb{R}^{d}\), with \(m=\E(h(\X))\), where \(\X\sim N({\bff 0},\V)\).  
If  \(i_{0},\dots,i_{n-1}\)  are independent random variables   uniformly distributed over \(\{1,\dots,d\}\), then  \begin{equation}
\label{eq:MSEUpper}
\E((\frac{\sum ^{n-1}_{j=0}h(\X_{j})}{n}-m)^{2})\leq 18\kappa^2\frac {d^{2}}{n}.
\end{equation}
\end{theorem}
\subsection{Tightness of MSE bound}
We now give an example where the bound on the mean square error in Theorem~\ref{th:MSE} is optimal, up to a multiplicative constant.  Let \(\V=\I\)  and let \(h(\x)=||\x||\) for \(\x\in\mathbb{R}^{d}\).
Thus  \(h\) is a \(1\)-Lipschitz function on
\(\mathbb{R}^{d}\). By~\cite[Sec. 11.3]{forbes2011statistical}, \begin{displaymath}
m= \frac{\sqrt2\Gamma(\frac{d+1}{2})}{\Gamma(\frac{d}{2})},
\end{displaymath}which implies by induction that  \(m\geq\sqrt{d/2}\). Furthermore, it follows from~\eqref{eq:xndefsimple} and by induction on \(n\) that \(X_{n}\)
has at most \(n\) non-zero components, and that the non-zero components of
 \(X_{n}\)
 are independent
standard Gaussian random variables. Thus  \(\E(||X_{n}||^{2})\leq n\), and
so  \(\E(||X_{n}||)\leq \sqrt{n}\) for \(n\geq 0\). Thus, \begin{displaymath}
\E(\sum ^{n-1}_{j=0}||\X_{j}||)\leq n^{3/2}.
\end{displaymath}Hence, for \(n=d/4\),\begin{equation*}E(m-\frac{\sum ^{n-1}_{j=0}h(\X_{j})}{n})\ge\frac{\sqrt{2}-1}{2}\sqrt{d}, \end{equation*}
and so \begin{equation*}
\E((\frac{\sum ^{n-1}_{j=0}h(\X_{j})}{n}-m)^{2})\ge \frac{d}{25}.
\end{equation*} Thus, the LHS of~\eqref{eq:MSEUpper}  is within an absolute constant from its RHS.
 
\section{The positive semi-definite case}\label{se:asymptotics}
Let \(h\) be a   \((\kappa,\gamma,\V)\)-Lipschitz function. Define \(m\), \(\Sigma\) and \(\hat h\) as in Section~\ref{se:Lipschitz}. This section assumes     that  \(V\) is positive definite and that, \(i_{n}\), \(n\geq0\),  are independent random variables   uniformly distributed over \(\{1,\dots,d\}\). Denote by    \(\lambda\) the smallest eigenvalue of \(\V\), and set 
\begin{equation*}\label{eq:defk'}
\kappa'=\frac{2\kappa d^{1+\gamma}}{\lambda\gamma}.
\end{equation*}  The following lemma, combined with Lemma~\ref{le:Wasserstein}, implies a
geometric bound on the Wasserstein distance between the distribution of \(\X_{n}\)
and \(N({\bff 0},\V)\) if \(\V\) is positive semi-definite.
\begin{lemma}\label{le:positiveSemi}For \(j\geq0\), \(\E(||\M_{j}||^{2})\leq d^{2}(1-\lambda d^{-1})^{j}\). 
\end{lemma}
 \commentt{\begin{proof}}{\proof{Proof.}}
We use the same notation as in the proof of Theorem~\ref{th:trSum}. Since the largest eigenvalue of \(\V\) is at most \(\tr(\V)=d\), it follows from~\eqref{eq:vjTVvj} that\begin{equation*}
\E(\vv^{T}\M_{j}^{T}\V\M_{j}\vv)\leq dE(||\vv_{j}||^{2}).
\end{equation*}On the other hand,~\eqref{eq:dvjTVvj} implies that 
\begin{equation*}
\E(||\vv_{j}||^{2})-\E(||\vv_{j+1}||^{2})\geq\lambda d^{-1}\E(||\vv_{j}||^{2}),
\end{equation*}and so \(\E(||\vv_{j}||^{2})\leq(1-\lambda d^{-1})^{j}\).
Hence, for any unit-vector \(\vv\), \begin{equation*}
\vv^{T}\E(\M_{j}^{T}\V\M_{j})\vv\leq d(1-\lambda d^{-1})^{j}.
\end{equation*} Thus each diagonal element of \(\E(\M_{j}^{T}\V\M_{j})\) is at most \(d(1-\lambda d^{-1})^{j}\), and so  \(\tr (\E(\M_{j}^{T}\V\M_{j}))\le d^{2}(1-\lambda d^{-1})^{j}\).   
This concludes the proof.
\commentt{\end{proof}}{\Halmos\endproof}
As noted before,    \( n^{-1}{\sum ^{n-1}_{j=0}h(\sqrt{\V}\Z_{j})}\)
is an unbiased estimator of \(m\). The
following lemma  implies that, if \(c>0\), the variance of this estimator is \(\Theta(n^{-1})\) as \(n\) goes to infinity. \begin{lemma}\label{le:asymptoticVariance}
As \(n\) goes to infinity, \(n^{-1} \E((\sum^{n-1}_{j=0}\hat h(\Z_{j}))^{2})\) converges to \(c\), where \begin{displaymath}
c=\var(h(\sqrt{\V}\Z_{0}))+2\sum_{j=1}^{\infty}\cov(h(\sqrt{\V}\Z_{0}),h(\sqrt{\V}\Z_{j})). \end{displaymath} \end{lemma}
Theorem~\ref{th:MSEAsymptot}
   below implies that,  if \(c>0\), \(\text{MSE}(n)\sim cn^{-1}\)  as \(n\) goes to infinity. 
    
\begin{theorem}\label{th:MSEAsymptot} As \(n\) goes to infinity,   \(n \E((\frac{\sum ^{n-1}_{j=0}h(\X_{j})}{n}-m)^{2})\)
  converges to \(c\) .
\end{theorem}
\commentt{\begin{proof}}{\proof{Proof.}} Define \(\beta_{j}\) and \(\beta\) via~\ref{eq:betajDef}, with \(b=0\). Let    \(\theta=(1-\lambda d^{-1})^{\gamma}\). By Lemma~\ref{le:ebarh2},  \begin{eqnarray*}\nonumber
\E(\beta_{j}^{2})
&\leq&\kappa^2 \E(||\M_{j}||^{2\gamma})\\
&\leq&\kappa^2 \E(||\M_{j}||^{2})^\gamma\\
&\le&\kappa^2 d^{2\gamma}\theta^{j}.\label{eq:Deltaj2}
\end{eqnarray*} The second inequality follows from Jensen's inequality, and
the last one from  Lemma~\ref{le:positiveSemi}. Hence, by the Cauchy-Schwartz inequality,  \begin{eqnarray*}\E(\beta^{2})&\leq&(\sum^{n-1}_{j=0}\theta^{j/2})\E(\sum^{n-1}_{j=0}\theta^{-j/2}\beta_{j}^{2})\\
&\le&\frac{\kappa^{2} d^{2\gamma}}{(1-\theta^{1/2})^{2}}\\
&\le&{\kappa'}^{2}.\end{eqnarray*}
The last inequality follows from  the relation  \(\theta ^{1/2}\leq1-\lambda \gamma d^{-1}/2\) (which is a consequence of Taylor's formula with Lagrange remainder). On the other hand,\begin{eqnarray*}\E((\sum^{n-1}_{j=0}\hat h(\Y_{j}))^{2})
&=&\E((-\beta+\sum^{n-1}_{j=0}\hat h(\Z_{j}))^{2})\\
&=&\E(\beta^{2})-2 \E(\beta\sum^{n-1}_{j=0}\hat h(\Z_{j}))+\E(( \sum^{n-1}_{j=0}\hat h(\Z_{j}))^{2}).
\end{eqnarray*}But, by the Cauchy-Schwartz inequality and  Lemma~\ref{le:asymptoticVariance},  for sufficiently large \(n\),
 \begin{eqnarray*}
|\E(\beta(\sum^{n-1}_{j=0}\hat h(\Z_{j})))|
&\leq&\kappa'\sqrt{\E\bigl((\sum^{n-1}_{j=0}\hat h(\Z_{j}))^{2}\bigr)}\\
&\le&\kappa'\sqrt{(c+1)n}.
\end{eqnarray*}
 Using Lemma~\ref{le:asymptoticVariance} once again, it follows that  \(n^{-1}\E((\sum^{n-1}_{j=0}\hat h(\Y_{j}))^{2})\) converges to \(c\) as \(n\) goes to infinity. This concludes the proof.
\commentt{\end{proof}}{\Halmos\endproof}We now show that the estimator \({(2/n)}{\sum ^{n-1}_{j=n/2}h(\X_{j})}\) of \(m\) has an exponentially decreasing bias. We also give a bound on the mean square error of this estimator which, in certain cases, is smaller than the RHS of~\eqref{eq:MSEUpper}.  
\begin{theorem}\label{th:MSEConditionning} Set \(\delta =\left\lceil 4(\lambda\gamma)^{-1}d\ln(\kappa d/\Sigma)  \right\rceil \). For \(d\geq3\) and even \(n>0\),  \begin{equation}
\label{eq:uppperBoundGeomBias}|\E(\frac{\sum ^{n-1}_{j=n/2}h(\X_{j})}{n/2}-m)|\le2\kappa' \frac{e^{-\lambda\gamma n/(4d)}}{n}.
\end{equation}Furthermore, if \(n>2\delta \),
\begin{equation}\label{eq:uppperBoundGeomMSE}
\E((\frac{\sum ^{n-1}_{j=n/2}h(\X_{j})}{n/2}-m)^{2})\le34\frac{\delta \Sigma^{2}}{ n}.
\end{equation}
\end{theorem}
\commentt{\begin{proof}}{\proof{Proof.}}  Set \(b=n/2\) and  define   \(\beta\)
via~\eqref{eq:betajDef}. Using calculations similar to the proof of  Theorem~\ref{th:MSEAsymptot}, it follows that \begin{displaymath}
\E(\beta^{2})\leq{\kappa'}^{2}(1-\lambda d^{-1})^{\gamma n/2}.  
\end{displaymath} Since  \(1+x\leq e^{x}\) for \(x\in\mathbb{R}\), it follows that\begin{displaymath}
|\E(\beta)|\leq{\kappa'}e^{-\lambda\gamma n/(4d)}. 
\end{displaymath} This implies~\eqref{eq:uppperBoundGeomBias} since the \(\hat h(\Z_{j})\)'s are centered.
 
  We now  prove~\eqref{eq:uppperBoundGeomMSE}.\ We first note that,  by applying~\eqref{eq:defkappaLips} with \(\X\sim N(\bff0,V)\),  \(\X'\sim N(\bff0,V)\), \(\X\) and \(\X'\) independent, if follows after some calculations that \begin{equation}\label{eq:SigKappad}
\Sigma^{2}\leq\kappa^{2}d,
\end{equation}and so \begin{equation}\label{eq:lowerBdelta}
\delta \geq  2(\lambda\gamma)^{-1}d.
\end{equation}   On the other
hand,
by Lemma~\eqref{le:positiveSemi} and Jensen's inequality,\begin{eqnarray*}\E(||\M_{j}||^{2\gamma})
&\leq& d^{2\gamma}(1-\lambda d^{-1})^{\gamma j}\\
&\le&d^{2\gamma}(1-\lambda\gamma d^{-1})^{ j}. \end{eqnarray*} The
second equation follows from the inequality \((1-\lambda d^{-1})^{\gamma}\leq 1-\lambda\gamma  d^{-1}\). Thus,     \begin{eqnarray*}
 \sum^{n-1}_{j=\delta }\E(||\M_{j}||^{2\gamma})
&\leq&\frac{d^{2\gamma+1}}{\lambda\gamma}(1-\lambda\gamma  d^{-1})^{\delta }\\ &\leq&\frac{d^{3}}{\lambda\gamma}\exp(-\lambda\gamma \delta d^{-1})\\ 
 &\le&\frac{d\Sigma^{2}}{\lambda\gamma\kappa^{2}}.
\end{eqnarray*}
Thus, by applying~Lemma~\ref{le:MSEGen},  it follows that\begin{displaymath}
\E((\frac{\sum ^{n-1}_{j=n/2}h(\X_{j})}{n/2}-m)^{2})\leq\frac{2}{n}(8\delta \Sigma^{2}+18\frac{d\Sigma^{2}}{\lambda\gamma} ).
\end{displaymath}
By~\eqref{eq:lowerBdelta}, this implies~\eqref{eq:uppperBoundGeomMSE}.
\commentt{\end{proof}}{\Halmos\endproof}
\section{Examples}\label{se:TheoExamples}
Let \(h\) be a     real-valued Borel function of \(d\) variables\comment{ that    $\E((h(\X))^{2})$ is finite,
for any Gaussian vector \(\X\)    with     \(\cov(\X)\le \V\).
We assume} that can be calculated at any point    in \(O(d)
\) time. Assume that \(\V\) is positive definite, and that both \(m=\E(h(\X))\) and \(\Sigma^{2}=\var(h(\X))\)
 exist and are finite, where   \(\X\sim N({\bff 0},\V)\).  Denote by MCMC the algorithm that generates  \(\X_{0},\dots,\X_{n-1}\) via~\eqref{eq:xndefsimple}, where the \(i_{j}\)'s are independent and identically distributed over \(\{1,\dots,d\}\), and estimates \(m\) via\begin{equation*}
h_{n,b}=\frac{\sum ^{n-1}_{j=b}h(\X_{j})}{n-b},
\end{equation*}  where \(b\) is a burn-in period. The standard Monte Carlo algorithm, referred to later as MC, first calculates a  lower-triangular matrix
  \(\A\)  satisfying~\eqref{eq:aat} in \(\Theta(d^{3})\) time via the procedure
described in~\cite[Subsection 2.3.3]{glasserman2004Monte}, then generates \(n'\) independent  \(d\)-dimensional vectors of independent standard Gaussian random variables 
\(\Z_{1},\dots,\Z_{n'}\), and estimates \(m\) by taking the average of  \(h(\A\Z_{j})\),
\(1\leq j\leq n'\). The
variance of this estimator is    \(V_{\text{MC}}(n')=\Sigma^{2}/n'\).  
\subsection{Comparison of the MC and MCMC methods}
\label{sub:compMC}The mean square error
of the \(h_{n,b}\) estimator of \(m\) is defined as   \begin{equation*}
\text{MSE}(n;b)=\E((h_{n,b}-m)^{2}).
\end{equation*}Given \(\epsilon\in(0,\Sigma)\),  \(n'=\Sigma^{2}/\epsilon^{2}\) samples of the MC algorithm are needed to ensure that     \(V_{\text{MC}}(n')=\epsilon^{2}\) (ignoring
rounding issues). Calculating the Cholesky decomposition and  \(h(\A\Z_{j})\),
\(1\leq j\leq n'\), takes  \begin{displaymath}
\tau_{\text{MC}}(\epsilon)=\Theta(d^{3}+\frac{\Sigma^{2}}{\epsilon^{2}}d^{2})
\end{displaymath}     
      time. On the other hand, for \(\epsilon>0\), if \(h\) is \((\kappa,\gamma,\V)\)-Lipschitz and \(\xi\in\{0,1/2\}\), denote
      by  \(\tau_{\text{MCMC}}(\epsilon,\xi)\) the running time of the MCMC algorithm needed  to ensure that  \(\text{MSE}(n;b)\le\epsilon^{2}\), using burn-in period \(b=\xi n\). If \(\gamma=1\), by Theorem~\ref{th:MSE}, after          \(n=18\kappa^{2}d^{2}/\epsilon^{2}\)   steps of the MCMC algorithm,   \(\text{MSE}(n)\le\epsilon^{2}\). Thus, $$\tau_{\text{MCMC}}(\epsilon,0)=O({\kappa^{2}}\frac{d^{3}}{\epsilon^{2}}).$$   Thus, if there is a constant \(\phi\geq1\)  independent of \(d\) such that  \(
\kappa^{2}d\leq\phi\Sigma^{2}\) (this is equivalent to saying that~\eqref{eq:SigKappad} is tight, up to
a constant),  then, for fixed \(\epsilon/\Sigma<1\),
    \begin{equation}\label{eq:tauMCMCfirst}
    \tau_{\text{MCMC}}(\epsilon,0)=O(d^{2}).
    \end{equation} Similarly, under the assumptions of Theorem~\ref{th:MSEConditionning}, 
\begin{equation*}
    \tau_{\text{MCMC}}(\epsilon,1/2)=O(\frac{\ln(\kappa d/\Sigma)d^{2}}{\lambda\gamma }\frac{\Sigma^{2}}{\epsilon^{2}}).
\end{equation*}   
Hence, if there are positive constants \(\phi\) and  \(\phi'\) independent of \(d\) such that  \(\kappa\le d^{\phi }\Sigma\) and  \(\lambda\gamma\geq\phi'\), then, for fixed \(\epsilon/\Sigma<1\),
 \begin{equation}
\label{eq:d2logd}
\tau_{\text{MCMC}}(\epsilon,1/2)=O(d^{2}\ln( d)).
\end{equation}
Examples where~\eqref{eq:tauMCMCfirst} or~\eqref{eq:d2logd} hold are given below. 
    
\subsection{A Basket option}
Consider  a set of \(d\) stocks \(S_{1},\dots,S_{d}\). For \(t\geq0\), denote by \(S_{i}(t)\) the price of \(S_{i}\) at time \(t\). Assume that \(S_{1}(0)=S_{2}(0)=\dots=S_{d}(0)=1\). A Basket call option with maturity \(T\) and strike \(K\) is a financial derivative that pays the amount
\(((S_{1}(T)+\cdots+ S_{d}(T))/d-K)^{+}\)
at time \(T\). Under a standard pricing model~\cite[Subsection 3.2.3]{glasserman2004Monte}, the price of a basket option is  \(\E(h(\U))\),
where \(\U\) is a centered Gaussian vector  with covariance matrix \(\V\) given by \(V_{ij}=\text{Correl}(\ln(S_{i}(T)),\ln(S_{j}(T)))\)
for \(1\leq i\leq j\leq d\), and, for \(x=(x_{1},\dots,x_{d})\in\mathbb{R}^{d}\), \begin{equation*}
h(\x)=(d^{-1}\sum^{d}_{i=1}\exp(-\frac{\sigma_{i} ^{2}}{2}T+\sigma_{i}\sqrt{T} \x_{i})-Ke^{-rT})^{+},
\end{equation*}
where \(r\) is the risk-free rate, and \(\sigma_{i}\) is the volatility of \(S_{i}\).
Assume that the \(\sigma_{i}\)'s are bounded by a constant independent of \(d\). It follows from Lemma~\ref{le:callLipschitz} below that \(h\) is \((\kappa ,\V)\)-Lipschitz, where \(\kappa=O(d^{-1/2})\)
as \(d\) goes to infinity.

\begin{lemma}\label{le:callLipschitz}Let \(g(x_{1},\dots,x_{d})=\max(\sum^{d}_{i=1}w_{i}e^{\sigma _{i}x_{i}}-K,0)\), where \(w_{i}\geq0\) for \(1\leq i\leq d\). Then  \(g\) is  \((\kappa,\V)\)-Lipschitz, where \(\kappa=\sqrt{\sum^{d}_{i=1}w_{i}^{2}e^{2\sigma_{i}^{2}}(4\sigma_{i}^{2}+1)}\).
\end{lemma}
 Thus, by Theorem~\ref{th:MSE}, \(n=O(d/\epsilon^{2})\) steps of the MCMC algorithm are sufficient  to ensure that  \(\text{MSE}(n)\le\epsilon^{2}\), and so   \(\tau_{\text{MCMC}}(\epsilon,0)=O(d^{2}/\epsilon^{2})\). On the other hand, if \(\Sigma=\Theta(1)\) (which  is the case~\cite[Sec.
25.14]{Hull12} if the volatilities and correlations are lower-bounded by a constant and \(K=0\), for instance),
then    \(\tau_{\text{MC}}(\epsilon)=\Theta(d^{3}+d^{2}/\epsilon^{2})\). In practice, though,  \(d\) is quite small.
\subsection{The multivariate normal function  }
Let \(a=(a_{1},\dots,a_{d})\in\mathbb{R}^{d}\), and   \(\hat a =\min_{1\leq i\leq d}|a_{i}|\). Set  \(h(\x)=1_{\x\le a}\) for \(\x=(x_{1},\dots,x_{d})\in\mathbb{R}^{d}\), where  \(\x\le a\) if and only if \(x_{i}\leq a_{i}\) for \(1\leq i\leq d\).  The following lemma and the analysis in Subsection~\ref{sub:compMC} show that, if there are positive constants \(\phi\) and  \(\phi'\) independent of \(d\)  such that  \(\hat a\ge d^{-\phi }\),  \(\Sigma\ge d^{-\phi }\), and   \(\lambda\geq\phi'\), then~\eqref{eq:d2logd} holds  for fixed \(\epsilon/\Sigma<1.\)\begin{lemma}\label{le:maxExample}  The function   \(h\)  is  \((3(d/\hat a)^{1/3},1/3,\bff W)\)-Lipschitz for any \(d\times d\) positive semi-definite matrix \(W\).   \end{lemma}
\commentt{\begin{proof}}{\proof{Proof.}}Let \(\nu>0\). It is easy to see that if \({||\X-\Y||}<\nu\) and, for \(1\leq i\leq d\), \(|X_{i}|\not\in[|a_{i}|,|a_{i}|+\nu]\) and \(|Y_{i}|\not\in[|a_{i}|,|a_{i}|+\nu]\),   then \(h(\X)=h(\Y)\). Hence
\begin{eqnarray*}
|h(\X)-h(\Y)|\le1_{\nu\leq||\X-\Y||}+\sum^{d}_{i=1}(1_{|X_{i}|\in[|a_{i}|,|a_{i}|+\nu]}+1_{|Y_{i}|\in[|a_{i}|,|a_{i}|+\nu]}).
\end{eqnarray*}
By Chebyshev's inequality,  \(\Pr(\nu\leq||\X-\Y||)\leq\nu^{-2}E(||\X-\Y||^{2}).\) On the other hand, a simple calculation shows that, for \(z>0\), the density of any centered Gaussian random variable at \(z\) is at most \(1/z\). Hence,
\begin{displaymath}
\Pr({|X_{i}|\in[|a_{i}|,|a_{i}|+\nu]})\leq 2\nu/\hat a,
\end{displaymath} and a similar relation holds for \(\Y_{i}\).  Thus, \begin{displaymath}
E(|h(\X)-h(\Y)|)\leq \nu^{-2}E(||\X-\Y||^{2})+4 d\nu/\hat a.
\end{displaymath}Minimizing over \(\nu\) implies that
 \begin{displaymath}
E(|h(\X)-h(\Y)|)\leq9(d/\hat a)^{2/3}(E(||\X-\Y||^{2}))^{1/3}.
\end{displaymath}
\commentt{\end{proof}}{\Halmos\endproof}

      \comment{Thus, by Theorem~\ref{th:MSE},\begin{displaymath}
\text{MSE}(n)\leq18\eta||\boldsymbol\alpha||^{2}\frac {d^{2}}{n}.
\end{displaymath}
}

\subsection{The maximum function}\label{sub:maxGen}
For \(x=(x_{1},\dots,x_{d})\in\mathbb{R}^{d}\), let \(h(x)=\max_{1\leq i\leq d}x_{i}\). Then  \(h\) is \(1\)-Lipschitz.  Let \(X\sim N(0,V)\), where \(V\) is a correlation matrix. Standard calculations show that \(\Pr(h(X)>z)\leq de^{-z^{2}/2}\) for \(z>0\). Thus, it follows after some calculations that \(\Pr(h(X)^{2}>z\ln (d))\leq e^{-z/4}\) for  \(z\geq2\) and  \(d\geq3\), and so  \(\E(h(X)^{2})\leq6\ln d\). On the other hand, since\begin{displaymath}
\Pr (X_{1}\in[4\sqrt{\ln d},5\sqrt{\ln d}])\geq\frac{d^{-15}}{\sqrt{2\pi}}, \end{displaymath}
where \(X_{1}\) is the first coordinate of \(X\), \begin{displaymath}
\Pr (h(X)\ge4\sqrt{\ln d})\geq\frac{d^{-15}}{\sqrt{2\pi}}. \end{displaymath}
Since \(E(h(X))\leq3\sqrt{\ln d}\), we conclude that \(\Sigma^{2}\geq d^{-15}/\sqrt{2\pi}\).
Hence~\eqref{eq:d2logd} holds  for fixed \(\epsilon/\Sigma<1\) if there is a positive constant  \(\phi'\) independent of \(d\) such that  \(\lambda \geq\phi'\).
\subsection{A  numerical example}
 \label{sub:NumericalExamples} In~\cite{gel2004}, the temperatures  \(Y({\bff s}_{1}),\cdots,Y({\bff s}_{d})\)   at a given set of \(d\) locations \({\bff s}_{1},\cdots,{\bff s}_{d}\)    in \(\mathbb{R}^{2}\) at a given
future time are modelled as a Gaussian vector   where \(\E(Y({\bff s}_{i}))\) is a known function of \({\bff s}_{i}\), \(\var(Y({\bff s}_{i}))=\varrho\) and, for two different locations \({\bff s}_{i}\) and \({\bff s}_{j}\),
\begin{equation*}
\cov(Y({\bff s}_{i}),Y({\bff s}_{j}))=\sigma ^{2}\exp(-\frac{||{\bff s}_{i}-{\bff s}_{j}||}{r}),
\end{equation*}where \(\varrho\), \(\sigma\) and \(r\) are positive constants, with \(\sigma^{2}<\varrho\).  By simulating
the vector   \((Y({\bff s}_{1}),\cdots,Y({\bff s}_{d}))\), we can  estimate the expected
maximum temperature
at these \(d\) locations.  

For simplicity of presentation, we assume thereafter that \(Y({\bff s}_{i})\) is centered for \(1\leq i\leq d\), and so  \(\X_{i}=\varrho ^{-1/2}Y({\bff s}_{i})\) is
 a standard  Gaussian random variable. The correlation matrix
 \(\V\)
 of the Gaussian vector \(\X=(X_{1},\dots,X_{d})^{T}\)  is given by 
 \begin{equation*}V_{ij}=\frac{\sigma ^{2}}{\varrho}\exp(-\frac{||{\bff s}_{i}-{\bff s}_{j}||}{r}),\end{equation*}  for \(i\neq j\).  Since the matrix \((\exp(-||{\bff s}_{i}-{\bff s}_{j}||/r))_{1\leq i,j\leq d}\) is positive semi-definite~\cite[Section 2.5]{cressie2015statistics}, \(\lambda\geq1-(\sigma ^{2}/\varrho)\). We use the MC and MCMC  algorithms to estimate   \(\E(\max_{1\le i\le d} \Y({\bff s_{i}})).
\) 
 Note that    \(\max_{1\le i\le d} \Y({\bff s_{i}})=h(X)
\), where  \(X=(X_1,\dots,X_d)\), and \(h(x)=\sqrt{\varrho}\max(x_{1},\dots,x_{d})\) for \(x=(x_{1},\dots,x_d)\in\mathbb{R}^{d}\). The analysis in Subsection~\ref{sub:maxGen} shows that~\eqref{eq:d2logd} holds  for fixed \(\epsilon/\Sigma<1\). 

Our numerical simulations assume that \(r=10\), \(\varrho=8\), \(\sigma^{2}=7.44\), and that, for \(1\leq i\leq d\),  the first (resp. second) coordinate of \({\bff s}_{i}\) equals \(\lfloor i/ d'\rfloor/ d'\) (resp. (\(i\mod d') / d') \), where \(d'=\lceil \sqrt{d}\rceil\).  After scaling, the  \(r\), \(\varrho\), and \(\sigma^{2}\) parameters are close to those estimated
 in~\cite{gel2004}. Our  experiments were performed on a desktop PC with
an Intel Pentium 2.90 GHz processor and 4 Go of RAM, running Windows 7 Professional. The
codes were written in the C++ programming language, and the compiler used was Microsoft Visual
C++ 2013. Computing times are given in seconds. Table~\ref{tab:Cholesky}  gives  computing times of the MC   method. Running the
 Cholesky factorization for \(d=10^{5}\) without external storage causes memory overflow.
 Extrapolating the results in  Table~\ref{tab:Cholesky}  shows that the Cholesky factorization would take a few weeks for \(d=10^{5}\)  if enough internal
 memory were available.  Table~\ref{tab:meteo} compares the MC and MCMC methods
 for \(d\) up to \(10^{4}\).   For the tested parameters,   the  MCMC  method is  more efficient than the MC method, and its efficiency   increases with \(d\).
For \(d=10^{5}\), \(n=100d\) and \(b=n/2\), the MCMC average is \(4.01\), and is calculated in \(8969\) seconds. Lemma~\ref{le:maxExample} shows that the MCMC method can also be used to calculate the probability that the maximum temperature over a set of \(d\) points exceeds a certain level.
\begin{table}[ht]
\centering
\caption{Running times           of  the MC
method. The second column gives
the   time to perform the Cholesky decomposition. The third column
gives the  time to simulate  
\(\Z_{1},\dots,\Z_{n'}\), and  calculating \(h(\A\Z_{i})\),
\(1\leq i\leq n'\), where \(n'=10^{4}\), and does not incorporate the Cholesky decomposition running time.
}
\label{tab:Cholesky}
\begin{tabular}{crr}
\\
\hline\hline
  $d$            &   Cholesky& Simulations        \\
               & decomposition &    \\\hline
10              & 0.000   & 0.007               \\
$10^2$      & 0.001   &   0.19                 \\
$10^3$          & 2.3  & 17.3                \\
$10^4$         & 2306   &            1702          
\end{tabular}
\end{table}
   
\begin{table}[ht]
\centering
\caption{The  MCMC and MC methods for estimating    \(\E(\max_{1\le i\le d} \Y({\bff s_{i}})),
\)
 with \(n=100d\) and \(\text{burnin}=n/2\).  The MCMC RMSE is an estimate of \(\sqrt{\text{MSE}(n;n/2)}\), which is calculated as explained in Section~\ref{se:estimatingMSE}. The standard deviation \(\Sigma\) is estimated using the MC method with \(n'=10^{4}\). The last column gives \(\tau_{\text{MC}}(\epsilon)/\tau_{\text{MCMC}}(\epsilon,1/2)\), where  \(\epsilon=\sqrt{\text{MSE}(n;n/2)}\), and \(\tau_{\text{MC}}(\epsilon)\) and \(\tau_{\text{MCMC}}(\epsilon,1/2)\)  are calculated in seconds.}
\label{tab:meteo}
\begin{tabular}{cccrcr}
&&&&&\\
\hline
$d$        & MCMC & MCMC  & MCMC &  \(\Sigma\)&$\tau_{\text{MC}}/\tau_{\text{MCMC}}$           \\
  &   Average       & RMSE &  comp. time        & &  \\
   \hline
\comment{10     &  & 0.32    & 0.034     & $0.0001$   &  & 0.005       &             & 1         }
$10^2$    &   2.38    & 0.119     & $0.002$     & 2.7&4          \\
$10^3$     & 3.02    & 0.069     & $0.19$     & 2.7&26              \\
$10^4$  &  3.57    & 0.049     & $16$  & 2.7       &173        \\
\end{tabular}
\end{table}
\subsection{Other examples}Spatial Gaussian processes of various  types  such as  Mat\'ern, powered exponential, and spherical, restricted   to \emph{any subset} of size \(d\) of \(\mathbb{R}^{2}\), are centered Gaussian vectors with a covariance
matrix   whose entries  can be calculated in  \(O(1)\) time. For instance, the   covariance
matrix \(V\) of a {\em powered  exponential} process restricted to a subset \(\{{\bff s}_{1},\dots,{\bff s}_{d}\}\) of  \(\mathbb{R}^{2}\) is given by~\cite{diggle2003introduction} 
 \begin{equation*}V_{ij}=\exp(-(\frac{||{\bff s}_{i}-{\bff s}_{j}||}{r})^{\theta}),\end{equation*}   where \(r>0\) and \(0<\theta\leq2\). The techniques of our paper can  be used to simulate the restriction of such processes to any finite subset of  \(\mathbb{R}^{2}\). \section{Conclusion}  We have shown how to simulate a Markov chain \(\X_{n}\), \(n\geq0\), such that the Wasserstein distance between the distribution of \(\X_{n}\) and \(N(\bff0,V)\) is at most \(d/\sqrt{n}\). It takes \(O(d)\) time   to generate each step of the chain.  Whereas  the standard Monte Carlo simulation method has \(\Theta(d^{2})\) storage cost, the storage cost of our method  is     \(\Theta (d)\).      Furthermore, by running the chain  \(n\) steps, our method can  estimate \(\E(h(\X))\), where \(\X\) is a centered Gaussian vector with covariance matrix \(\V\) and \(h\) is a real-valued function of \(d\) variables. Under certain conditions, we give an explicit upper bound on the mean square error of our estimate, and  show that it is inversely proportional to the running time. We  also prove that, in certain cases, the total  time needed by our method to obtain a given  standarized mean square error is \({O}^*(d^{2})\) time, whereas the standard Monte Carlo method takes  \(\Theta(d^{3})\)  time. \comment{Our algorithms do not require the knowledge of a lower bound on the smallest eigenvalue of \(V\), but the analysis in Section~\ref{se:asymptotics} and part of Section~\ref{se:TheoExamples} does.}


\comment{
\begin{lemma}
Let \(\lambda\) be the smallest eigenvalue
of \(\V\). Then \(
||\M^{n}\f_{i}||\leq e^{-\lambda^{2}n/(28d^{2})}\).
\end{lemma}
 \commentt{\begin{proof}}{\proof{Proof.}}
The second part of the lemma clearly holds if \(\lambda=0\). Assume now that
\(\lambda>0\) and let \(N=\M^{T}\M\). By applying the first part of the lemma to the sequence of vectors  \(\f_{d},\ldots,\f_{1}, \f_{1},\ldots,\f_{d}\), we conclude that
\(
||N^{n}\f_{i}||\leq (2d/n)^{1/2}
\) for \(n\geq1\) and \(1\leq i\leq d\). Let \(\mu\) be the largest eigenvalue
of \(N\) and
\(\vv\) be a corresponding unit-length eigenvector. Then \begin{equation}
\V^{-1}\vv=\sum^{d}_{i=1}\boldsymbol\alpha_{i}\e_{i}
\end{equation}
and so \begin{equation}
\vv=\sum^{d}_{i=1}\boldsymbol\alpha_{i}\f_{i}.
\end{equation}
Consequently,
\begin{equation}
\mu^{n}\vv=\sum^{d}_{i=1}\boldsymbol\alpha_{i}N^{n}\f_{i}.
\end{equation}
Hence\begin{equation}
\mu^{n}\le(2d/n)^{1/2}\sum^{d}_{i=1}|\boldsymbol\alpha_{i}|.
\end{equation}
Since \(||\V^{-1}\vv||\leq\lambda^{-1}\), we conclude using the Cauchy-Schwartz inequality that    
\begin{equation}\label{eq:mu^n}
\mu^{n}\le d\lambda^{-1}(2/n)^{1/2}.
\end{equation}
By applying Eq.~\ref{eq:mu^n} to the case where \(n\) is the ceil of \(2e(d/\lambda)^{2}\),
it follows that \(
\mu\le e^{-\lambda^{2}/(14d^{2})}
\).
The second part of the lemma follows by noting that \(||\M||^{2}=\mu\).
\commentt{\end{proof}}{\Halmos\endproof}
}
\comment{
Assume now that \(\vv\) is an eigenvector of \(\V\) with eigenvalue \(\lambda\).
By the Cauchy-Schwartz inequality, for \(0\leq i\leq d-1\),
\begin{equation}
(\f_{i+1}^{T}\vv_{i})^{2}-(\f_{i+1}^{T}\vv)^{2}\ge-2||\vv-\vv_{i}||.
\end{equation}
Since
\begin{equation}
||\vv_{i}-\vv_{i+1}||^{2}=(\f_{i+1}^{T}\vv_{i})^{2},
\end{equation} 
we conclude, using Eq., that\begin{equation}
||\vv_{i+1}||^{2}\leq||\vv_{i}||^{2}+2||\vv-\vv_{i}||-(\f_{i+1}^{T}\vv)^{2}.
\end{equation}
Since  \(\sum^{d}_{i=1}(\f_{i}^{T}\vv)^{2}=\vv^{T}Vv\), it follows that 
\begin{equation}
||\vv_{d}||^{2}\leq1+2\sum^{d-1}_{i=0}||\vv-\vv_{i}||-\lambda
\end{equation}On the other hand, 
  \begin{equation*}||\vv||^{2}&=&||\vv_{i}||^{2}+\sum^{i-1}_{j=0}||\vv_{j}-\vv_{j+1}||^{2}\end{equation*}
  and so\begin{equation}
||\vv-\vv_{i}||^{2}\leq i(1-||\vv_{d}||^{2}).
\end{equation}
If there is \(i\in\{1,\ldots,d\}\) such that \(||\vv-\vv_{i}||\geq\boldsymbol\alpha\sqrt{i}\)
}
\comment{
 and so, by the Cauchy-Schwartz inequality, 
\begin{equation}
||\vv_{d}||^{2}+\lambda-1\leq \sqrt{2d(d-1)}\sqrt{\sum^{d-1}_{i=1}||\vv-\vv_{i}||^{2}/i}
\end{equation}We conclude that
 \begin{equation}
||\vv_{d}||^{2}+\lambda-1\leq (d-1)\sqrt{2d}\sqrt{1-||\vv_{d}||^{2}}
\end{equation}
}
\comment{
\begin{lemma}
\begin{equation}
\tr(\A\B)\leq||\A||_{2}||\B||
\end{equation}
\end{lemma}
 \commentt{\begin{proof}}{\proof{Proof.}}
Let \(\Pf\) and \(Q\) be two orthogonal matrices such that \(\B=PDQ\), where
\(D\) is a diagonal matrix containing the singular values of \(\B\). Then
\begin{equation}
\tr(\A\B)=\tr(\A'D),
\end{equation}
where \(\A'=Q\A\Pf\). Since \(\A\) and \(\A'\) have the same singular values, \(||\A||_{2}=||\A'||_{2}\).
Thus all diagonal elements of  \(\A'\) are upper-bounded by  \(||\A||_{2}\),
and so \(\tr(\A'D)\leq||\A||_{2}\tr(D)\). But \(\tr(D)=||\B||\).
\commentt{\end{proof}}{\Halmos\endproof}
}

\commentt{\appendix}
{\begin{APPENDIX}{}}
 \section{Proof of Lemma~\ref{le:deterministic}}
 Recall first that if \(\Z\) and \(\Z'\) are independent centered  \(d\)-dimensional random vectors such that \(\E(||\Z||^{2})\) and \(\E(||\Z'||^{2})\)  are finite, then  \(\E(||\Z||^{2})=\tr(\cov(\Z))\),      \(\cov(\A\Z)=\A\cov(\Z)\A^{T}\), and \(\cov(\Z+\Z')=\cov(\Z)+\cov(\Z')\). 

It can be shown by induction that  \(\Y_{n}\) is a linear combination of 
   \(g_{{0}},\dots,g_{{n-1}}\), and so \(\Y_{n}\) is  a centered Gaussian vector. Hence   \(\X_{n}\) is also centered and Gaussian. Furthermore,  \(\Z_{n}\) is centered Gaussian since it is a linear combination of \(g_{{0}},\dots,g_{{n-1}}\) and of \(\Z_{0}\). Thus, \(\Z_{n}\) and \(g_{n}\) are independent. For \(0\le  l\leq n\),\begin{eqnarray}\Z_{l+1}&=&\Y_{l+1}+\M_{l+1}\Z_{0}\nonumber
 \\&=&\Pf_{i_{l}}\Y_{l}+g_{l}\f_{i_{l}}+\Pf_{i_{l}}\M_{l}\Z_{0}\nonumber
\label{eq:Znrecursion}\\&=&\Pf_{i_{l}}\Z_{l}+g_{l}\f_{i_{l}}.
  \end{eqnarray}Thus, since  \(\Z_{l}\) and \(g_{l}\) are independent, 
  \begin{eqnarray*}\cov(\Z_{l+1})
  &=&\cov(\Pf_{i_{l}}\Z_{l})+\cov(g_{l}\f_{i_{l}})\\
  &=&\Pf_{i_{l}}\cov(\Z_{l}){P}^{T}_{i_{l}}+\E(g_{l}^{2})\f_{i_{l}}\f_{i_{l}}^{T}\\
  &=&\Pf_{i_{l}}\cov(\Z_{l})\Pf_{i_{l}}+\f_{i_{l}}\f_{i_{l}}^{T}.
  \end{eqnarray*}      
It follows by induction that \(\cov(\Z_{l})=\I\), and so \(\Z_{l}\sim N(\bff0,\I)\). Thus, \eqref{eq:covZ} holds when \(m=n\). Furthermore, since \(g_{l}\) and \(\Z_{m}\) are independent  for \(0\le m\le l\le\ n-1\), it follows from~\eqref{eq:Znrecursion} that \(\E(\Z_{l+1}\Z^{T}_{m})=\Pf_{i_{l}}\E(\Z_{l}\Z^{T}_{m})   \). It follows by induction on \(l\) that \(\E(\Z_{l}\Z^{T}_{m})=\M_{m,l}\) for \(0\le m\le l\le\ n\), Hence~\eqref{eq:covZ}.

 Since  \(\Y_{n}\) is a linear combination of 
   \(g_{{0}},\dots,g_{{n-1}}\), the vectors \(\Z_{0}\) and \(\Y_{n}\) are independent.
   Thus, as \(\cov(\Z_{n})=\I\) and    \(\cov(\M_{n}\Z_{0})=\M_{n}\M_{n}^{T}\),
   it follows from \eqref{eq:wWn+1} that \begin{equation*}\label{eq:covYn}
\cov(\Y_{n})=\I-\M_{n}\M_{n}^{T}.
\end{equation*}  Hence~\eqref{eq:rn}, which implies that  \(\cov(\X_{n})\leq \V\). On the other hand, by \eqref{eq:wWn+1}, \(\sqrt{\V}\Z_{n}-\X_{n}=\sqrt{\V}\M_{n}\Z_{0}\) and so,  $$\sqrt{\V}\Z_{n}-\X_{n}\sim N({\bff 0},\sqrt{\V}\M_{n}\M_{n}^{T}\sqrt{\V}).$$ Hence,\begin{eqnarray*}\E(||\X_{n}-\sqrt{\V}\Z_{n}||^{2})
&=&\tr(\sqrt{\V}\M_{n}\M_{n}^{T}\sqrt{\V})\nonumber\\
\label{eq:XnVZn}&=&||\M_{n}||^{2}.
\end{eqnarray*} The second equation follows from the relation   \(\tr(\A\B)=\tr(\B\A)\). Using~\eqref{eq:rn}, this implies~\eqref{eq:trRn}. This concludes the proof.  \commentt{\qed}{\Halmos\endproof}
 \section{Proof of Lemma~\ref{pr:GaussianLip1}}
Let \((X,X')\) be a Gaussian vector in \(\mathbb{R}^{2}\), with \(X\sim N({ 0},\nu)\), \(X'\sim N( 0,\nu')\) and \(\nu'\leq\nu\). We show the following, which immediately implies Lemma~\ref{pr:GaussianLip1}:
 \begin{equation}\label{eq:expLipsch}
\E((e^{X}-e^{X'})^{2})\leq (\nu +\nu '+1/2)(e^{2\nu}+e^{2\nu'})\E((X-X')^{2}).
\end{equation}Let \begin{displaymath}
\rho=\E((X-X')^{2})=\nu +\nu '-2\cov(X,X').
\end{displaymath}  Since \(\E(e^{Z})=e^{\frac{1}{2}\var (\Z)}\) for any centered Gaussian random variable \(Z\), \begin{eqnarray*}
\E((e^{X}-e^{X'})^{2})&=&\E(e^{2X}+e^{2X'}-2e^{X+X'})\\
&=&e^{2\nu}+e^{2\nu'}-2e^{\frac{1}{2}\var(X+X')}\\
&=&e^{2\nu}+e^{2\nu'}-2e^{\nu/2+\nu'/2+\cov(X,X')}\\
&=&2e^{\nu+\nu'}(\cosh(\nu-\nu')-e^{-\rho/2})\\&\le&e^{\nu+\nu'}((\nu-\nu')^{2}\cosh(\nu-\nu')+\rho).\end{eqnarray*}The last equation follows from the inequalities \(1-x\leq e^{-x}\) and \(\cosh(x)\leq1+x^{2}\cosh(x)/2\) (which is a consequence of Taylor's formula with Lagrange remainder) for any real number \(x\).
On the other hand,  \((\sqrt\nu-\sqrt{\nu'})^{2}\le\rho\) since \(\cov(X,X')\leq\sqrt{\nu\nu'}\), and so \begin{eqnarray}\label{eq:nunu'}(\nu-\nu')^{2}\nonumber
&=&(\sqrt\nu+\sqrt{\nu'})^{2}(\sqrt\nu-\sqrt{\nu'})^{2}\\\nonumber
&\le&\rho(\sqrt\nu+\sqrt{\nu'})^{2}\\
&\le&2\rho(\nu+\nu').
 \end{eqnarray}Hence, as \(1\leq\cosh(x)\) for any real number \(x\),
 \begin{eqnarray*}\E((e^{X}-e^{X'})^{2})
 &\le&\rho e^{\nu+\nu'}(2(\nu+\nu')\cosh(\nu-\nu')+1)\\
 &\le&\rho e^{\nu+\nu'}\cosh(\nu-\nu')(2\nu+2\nu'+1).
 \end{eqnarray*}This implies~\eqref{eq:expLipsch}.

\section{Proof of Lemma~\ref{le:barhZj}}
We first prove the following lemma which implies that, under certain conditions, the   covariances between the components of \(\Y\) and \(\Y'\) can be used to bound the covariance between \(\hat h_{1}(\Y)\) and \(\hat h_{2}(\Y')\).

\begin{lemma}\label{le:CovLipsh}
Let \(\begin{pmatrix}\Y \\
\Y' \\
\end{pmatrix}\) be a  Gaussian column vector in \(\mathbb{R}^{2d}\), with \(\Y\sim \Y'  \sim N({\bff0},\I)\)  and \(\E(\Y\Y'^{T})=\A\B^{T}\), where \(\A\) and \(\B\) are \(d\times d\) matrices,
with \(\A\A^{T}\leq\I\) and \(\B\B^{T}\leq\I\). Let \(h_{1}\) and \(h_{2}\) be two \((\kappa,\gamma,\V)\)-Lipschitz functions on \(\mathbb{R}^{d}\). Then   \begin{equation*}
|\E(\hat h_{1}(\Y)\hat h_{2}(\Y'))|\leq \kappa^2(||\A||^{2\gamma}+||\B||^{2\gamma}).
\end{equation*}   
\end{lemma}
 \commentt{\begin{proof}}{\proof{Proof.}}
We first note that, if  \(\Z\sim\Z'\sim N({\bff0},\I)\) and \(\Z\), \(\Z'\)
are independent, then   \begin{equation}\label{eq:barhZZ'}
\E(\hat h_{1}(\Z)\hat h_{2}(\Z'))=0.
\end{equation} Furthermore, if \((\Y,\Y')\) is a centered Gaussian vector in \(\mathbb{R}^{2d}\) with      \(\cov(\Y)\le \I\) and
  \(\cov(\Y')\le \I\), and  \(h_{}\) is a  \((\kappa,\gamma,\V)\)-Lipschitz function on \(\mathbb{R}^{d}\), then \begin{eqnarray}\E((\hat h(\Y)-\hat h(\Y'))^{2})&\leq& \kappa^2(\E(||\sqrt{\V}\Y-\sqrt{\V}\Y'||^{2}))^{\gamma}\nonumber
\\ 
&=&\kappa^2(\tr(\sqrt{\V}\cov(\Y-\Y')\sqrt{\V}))^{\gamma}
\nonumber
\\ 
&=&\kappa^2(\tr(\V\cov(\Y-\Y')))^{\gamma}
.\label{eq:barhCov}\end{eqnarray} The second equation follows from the fact that \(\sqrt{\V}(\Y-\Y')\) is a centered Gaussian vector with covariance matrix \(\sqrt{\V}\cov(\Y-\Y')\sqrt{\V}\).

We now prove the lemma.  Let   \(\bff G\), \({\bff G'}\), \({\bff G''}\),  \({\bff G}_{1}\) and  \({\bff G}_{2}\) be independent \(d\)-dimensional  Gaussian vectors such that \({\bff G}\sim {\bff G'} \sim {\bff G''} \sim N({\bff0},\I)\), \({\bff G}_{1}\sim N({\bff0},\I-\A\A^{T})\), and  \({\bff G}_{2}\sim N({\bff0},\I-\B\B^{T})\). Note that \({\bff G}_{1}\) and \({\bff G}_{2}\) exist since \(\I-\A\A^{T}\) and   \(\I-\B\B^{T}\) are  positive semi-definite. Since  \({\bff {\bff AG}}\sim N({\bff0},\A\A^{T})\) and is independent of \({\bff G}_{1}\), \({\bff {\bff AG}}+{\bff G}_{1}\sim N({\bff0},\I)\). Similarly, \({\bff BG}+{\bff G}_{2}\sim N({\bff0},\I)\). Also, since \({\bff G}\), \({\bff G}_{1}\) and \({\bff G}_{2}\) are independent and centered,\begin{eqnarray*}\E(({\bff {\bff AG}}+{\bff G}_{1})({\bff BG}+{\bff G}_{2})^{T})&=& \E(({\bff {\bff AG}})({\bff BG})^{T})\\&=&\A\B^{T}.\end{eqnarray*} Thus, the covariance matrix of the Gaussian vector \(\begin{pmatrix}{\bff {\bff AG}}+{\bff G}_{1} \\
{\bff BG}+{\bff G}_{2} \\
\end{pmatrix}\)  is \(\left(\begin{array}{cc}
\I & \A\B^{T} \\
\B\A^{T} & \I \\
\end{array}\right)\). Hence the centered Gaussian vectors  \(\begin{pmatrix}{\bff {\bff AG}}+{\bff G}_{1} \\
{\bff BG}+{\bff G}_{2} \\
\end{pmatrix}\)  and \(\begin{pmatrix}\Y \\
\Y' \\
\end{pmatrix}\)     have the same covariance matrix, and so they have the same distribution.  Thus,\begin{eqnarray*}\E(\hat h_{1}(\Y)\hat h_{2}(\Y'))
&=&\E(\hat h_{1}({\bff {\bff AG}}+{\bff G}_{1})\hat h_{2}({\bff BG}+{\bff G}_{2}))\\
&=&\E((\hat h_{1}({\bff {\bff AG}}+{\bff G}_{1})-\hat h_{1}({\bff {\bff AG}}'+{\bff G}_{1}))(\hat h_{2}({\bff BG}+{\bff G}_{2})-\hat h_{2}({\bff BG}''+{\bff G}_{2}))\\
&\le&\frac{1}{2}(\E((\hat h_{1}({\bff {\bff AG}}+{\bff G}_{1})-\hat h_{1}({\bff {\bff AG}}'+{\bff G}_{1}))^{2})+\\&&\hskip14em \E((\hat h_{2}({\bff BG}+{\bff G}_{2})-\hat h_{2}({\bff BG}''+{\bff G}_{2}))^{2}))
\\&\le&\kappa^2(( \tr(\V\A\A^{T}))^{\gamma}+( \tr(\V\B\B^{T}))^{\gamma}).
\\&=&\kappa^2(( \tr(\A^{T}\V\A))^{\gamma}+( \tr(\B^{T}\V\B))^{\gamma}).
\end{eqnarray*}
The second equation follows by applying~\eqref{eq:barhZZ'} to each of the pairs \(({\bff {\bff AG}}+{\bff G}_{1},{\bff BG}''+{\bff G}_{2})\),  \(({\bff {\bff AG}}'+{\bff G}_{1},{\bff BG}''+{\bff G}_{2})\), and  \(({\bff {\bff AG}}'+{\bff G}_{1},{\bff BG}+{\bff G}_{2})\). The fourth equation follows from~\eqref{eq:barhCov} and the relations \(\cov({\bff {\bff AG}}-{\bff {\bff AG}}')=2\A\A^{T}\) and  \(\cov({\bff BG}-{\bff BG}'')=2\B\B^{T}\). 

Hence\begin{equation*}
\E(\hat h_{1}(\Y)\hat h_{2}(\Y'))\leq \kappa^2(||\A||^{2\gamma}+||\B||^{2\gamma}).
\end{equation*}   
Replacing \(h_{2}\) with \(-h_{2}\) concludes the proof.\commentt{\end{proof}}{\Halmos\endproof}
  
We now prove Lemma~\ref{le:barhZj}.
Assume first that the sequence \(i_{0},\dots,i_{n-1}\), is deterministic.
By Lemma~\ref{le:deterministic},  \(\Z_{j}\sim \Z_{l}\sim N({\bff0},\I)\) for \(0\leq j\leq l\leq n\), and  \begin{equation*}
\E(\Z_{l}\Z^{T}_{j})=\M_{j,l}.
\end{equation*}
Let \(\A=\M_{j',l}\) and  \(\B=\M_{j,j'}^{T}\), with \(j'=\lfloor (j+l)/2\rfloor\).
Since \(\A^T\) is the product of \(l-j'\) projection matrices, it follows
by induction on \(l\) that \(||\A^{T}\x||\leq||\x||\) for \(x\in\mathbb{R}^{d}\), and so  \(\A\A^{T}\leq\I\).     Similarly,  \(\B\B^{T}\leq\I\). Since \(\M_{j,l}=\A\B^{T}\),  it follows from Lemma~\ref{le:CovLipsh} that\begin{equation}\label{le:covHHat}|\E(\hat h(\Z_{j})\hat h(\Z_{l}))
|\le \kappa^2(||\M_{j',l}||^{2\gamma}+||\M_{j,j'}^{T}||^{2\gamma}).
\end{equation}Thus,  
\begin{eqnarray}
\label{eq:varianceUnbiasedDet}
\E((\sum^{n-1}_{j=b}\hat h(\Z_{j}))^{2})
&\le&2\sum^{n-1}_{ j=b}\sum_{l=j}^{n-1}\E(\hat h(\Z_{j})\hat h(\Z_{l}))\nonumber\\
&=&2\sum _{b\leq j\leq l\leq n-1,\,l-j< 2\delta }\E(\hat h(\Z_{j})\hat h(\Z_{l}))+2\sum _{b\leq j\leq l\leq n-1,\,l-j\geq 2\delta }\E(\hat h(\Z_{j})\hat h(\Z_{l}))\nonumber\\
&\le&4(n-b)\delta \Sigma^{2}+2\kappa^2\sum_{b\leq j\leq l\leq n-1,\,l-j\geq 2\delta }^{n-1}||\M_{j',l}||^{2\gamma}+||\M_{j,j'}^{T}||^{2\gamma}\nonumber\\
&\le&4(n-b)\delta \Sigma^{2}+4\kappa^2\sum_{b\leq j\leq l\leq n-1,\,l-j\geq \delta }^{n-1}||\M_{j,l}||^{2\gamma}+||\M_{j,l}^{T}||^{2\gamma}.\end{eqnarray}The third equation follows from observing that  \(\hat h(\Z_{j})\) and   \(\hat h(\Z_{l})\) are centered and have standard deviation \(\Sigma\). The last equation follows from the fact that each term \(\M_{j',l}\) (resp. \(\M_{j,j'}\)) occurs  at most twice in the last line. 

Assume now that the sequence \(i_{0},\dots,i_{n-1}\), is random. By~\eqref{eq:varianceUnbiasedDet},
    \begin{equation*}
\E((\sum^{n-1}_{j=b}\hat h(\Z_{j}))^{2}|i_{0},\dots,i_{n-1})\leq 4(n-b)\delta \Sigma^{2}+4\kappa^2\sum _{b\leq j\leq l\leq n-1,\,l-j\geq \delta }||\M_{j,l}||^{2\gamma}+||\M_{j,l}^{T}||^{2\gamma}.
\end{equation*} We conclude the proof by taking expectations and using the tower law.
  \commentt{\qed}{\Halmos\endproof}
   
\section{Proof of Lemma~\ref{le:asymptoticVariance}}
Let \(a_{0}=\E((\hat h(\Z_{0}))^{2})\) and, for \(j>0\), let \(a_{j}=2\E(\hat h(\Z_{0})\hat h(\Z_{j}))\). Since \((\Z_{j})\), \(j\geq0\), is a time-homogeneous Markov Chain and \(\Z_{j}\sim N(\bff0,I)\), \(2\E(\hat h(\Z_{k})\hat h(\Z_{k+j}))=a_{j}\) for \(j>0\). Hence \begin{eqnarray*} 
\E((\sum^{n-1}_{j=0}\hat h(\Z_{j}))^{2})
&=&\sum^{n-1}_{j=0}\E((\hat h(\Z_{j})^{2})+2\sum_{0\leq k<k+j<n} \E(\hat h(\Z_{k})\hat h(\Z_{k+j}))\\  
&=&na_{0}+\sum_{0\leq k<k+j<n} a_{j},\end{eqnarray*} and so\begin{equation}
\label{eq:asymtoticBarhZ}n^{-1} \E(\sum^{n-1}_{j=0}\hat h(\Z_{j}))^{2}=\sum^{n-1}_{j=0}\frac{n-j}{n}a_{j}.  \end{equation}On the other hand, by applying~\eqref{le:covHHat} with \(j=0\) and \(l=j\),
it follows that
\begin{eqnarray*}|\E(\hat h(\Z_{0})\hat h(\Z_{j}))
|&\le& \kappa^2 \E(||\M_{j',j}||^{2\gamma}+||\M_{j'}^{T}||^{2\gamma})\\
&\le& \kappa^2 ((\E(||\M_{j',j}||^{2}))^{\gamma}+(\E(||\M_{j'}^{T}||^{2}))^{\gamma}),
\end{eqnarray*} with \(j'=\lfloor j/2\rfloor\). The second equation follows from Jensen's inequality. But, since \(\M_{j',j}\sim  \M_{j-j'}\),  \begin{equation*} \E(||\M_{j',j}||^{2})=  \E(||\M_{j-j'}||^{2}).
\end{equation*}Furthermore, as \(j'\leq j-j'\), it follows from Theorem~\ref{th:trSum}
that  \begin{equation*}\E(||\M_{j-j'}||^{2})
\leq \E(||\M_{j'}||^{2}).
\end{equation*} Since \(\M_{j'}\sim \M_{j'}^{T}\), we conclude that \(|a_{j}
|\leq 2\kappa^2 (\E(||\M_{j'}||^{2}))^{\gamma}.
\) Hence, by Lemma~\ref{le:positiveSemi},  the series \(\sum_{j=0}^\infty a_{j}\) is absolutely convergent. Thus, the RHS of~\eqref{eq:asymtoticBarhZ} converges to \(\sum^{\infty}_{j=0}a_{j}\)  as \(n\) goes to infinity. But \(\E(\hat h(\Z_{0})\hat h(\Z_{j}))=\cov(h(\sqrt{\V}\Z_{0}),h(\sqrt{\V}\Z_{j}))\) since \(\E(h(\sqrt{\V}\Z_{0}))=\E(h(\sqrt{\V}\Z_{j}))=m\), and so  \(\sum^{\infty}_{j=0}a_{j}=c\). This concludes the proof.\commentt{\qed}{\Halmos\endproof}
\section{Proof of Lemma~\ref{le:callLipschitz}}Since  the function \(\max(z-K,0)\) is \(1\)-Lipschitz with respect to \(z\), we can assume without loss of generality that \(K=0\). Let  \(\begin{pmatrix}\U \\
\U' \\
\end{pmatrix}\) be a centered Gaussian vector  with     \(\cov(\U)\le \V\) and
  \(\cov(\U')\le \V\), where \(\U\) and \(\U'\) have dimension \(d\).  Let \(U_{i}\) (resp. \(U'_{i}\)) be the \(i\)-th component of \(\U\) (resp. \(\U'\)) and \(\lambda _{i}=(4\sigma_{i}^{2}+1)e^{2\sigma_{i}^{2}}\). It follows from  Lemma~\ref{pr:GaussianLip1} that \begin{displaymath}
\E((e^{\sigma _{i}U_{i}}-e^{\sigma _{i}U'_{i}})^{2})\leq\lambda _{i}\E((U_{i}-U_{i}')^{2}).
\end{displaymath}On the other hand, by the Cauchy-Schwartz inequality, \begin{eqnarray*}
(g(\U)-g(\U'))^{2}&=&\left( \sum^{d}_{i=1}(\lambda _{i}^{1/2}w_{i})(\frac{e^{\sigma _{i}U_{i}}-e^{\sigma _{i}U'_{i}}}{\lambda _{i}^{1/2}})\right)^{2}\\&\leq& (\sum^{d}_{i=1}\lambda _{i}w_{i}^{2})\sum^{d}_{i=1}\frac{(e^{\sigma _{i}U_{i}}-e^{\sigma _{i}U'_{i}})^{2}}{\lambda _{i}}.
\end{eqnarray*} Taking expectations, it follows that \begin{eqnarray*}\E((g(\U)-g(\U'))^{2})&\leq& (\sum^{d}_{i=1}\lambda _{i}w_{i}^{2})(\sum^{d}_{i=1}\E((U_{i}-U'_{i})^{2}))
\\&\\&=&\kappa^2 \E(||\U-\U'||^{2}),\end{eqnarray*}as desired.
 \commentt{\qed}{\Halmos\endproof}

\section{Estimating the mean square error}
\label{se:estimatingMSE}This section describes a numerical
method to estimate the mean square error of the MCMC method for moderate
values of \(d\). 
  The method first calculates   a matrix  \(\A\) satisfying~\eqref{eq:aat} using Cholesky factorization. An unbiased  estimator of \(m\) is then computed as follows. Define the Markov chain \((\X'_{n})\), \(n\ge0\),
by setting  \(\X'_{0}=\A\Z_{0}\)   and, for   \(n\geq0\),  let
 \begin{equation}\label{eq:xndefsimpleUB}
\X'_{n+1}=\X'_{n}+(g_{n}-\e_{i_{n}}^{T}\X'_{n})(\V \e_{i_{n}}).
\end{equation}Thus \(\X'_{n}\) satisfies the same recursion as \(\X_n\).
By rewriting~\eqref{eq:xndefsimpleUB} as   \begin{equation*}
\X'_{n+1}=(\I-\V\e_{i_{n}}\e_{i_{n}}^{T})\X'_{n}+g_{n}(\V\e_{i_{n}}),
\end{equation*}
and using calculations similar to those in the proof of Lemma~\ref{le:deterministic},
it can be shown by induction that \(\X'_{n}\sim N(\bff0,V)\). Thus, \begin{equation*}
h'_{n,b}=\frac{\sum ^{n-1}_{j=b}h(\X'_{j})}{n-b}
\end{equation*}
is an unbiased estimator of \(m\).  Since \(\text{MSE}(n;b)\) equals the
variance of \(h_{n,b}\) plus its square bias,  \begin{equation}\label{eq:MSEestimate}
\text{MSE}(n;b)=\var(h_{n,b})+(\E(h_{n,b}-h'_{n,b}))^{2}.
\end{equation}
 In Subsection~\ref{sub:NumericalExamples}, 
 \(\text{MSE}(n;b)\) is estimated for \(d\leq10^{4}\)    using~\eqref{eq:MSEestimate},
 each term in the RHS of~\eqref{eq:MSEestimate} being calculated via  \(100\) independent simulations of    \(h_{n,b}\)
and \(h'_{n,b}\). For any \(j\in\{0,\dots,n-1\}\), the same random variables \(g_{j}\)
and \(i_{j}\) are used to calculate  \(h_{n,b}\)
and \(h'_{n,b}\) via~\eqref{eq:xndefsimple} and~\eqref{eq:xndefsimpleUB}. 
\commentt{}{\end{APPENDIX}}
\section{Acknowledgments} This research has been presented at the Paris Bachelier Seminar, November 2016,
and at the  2nd IMA Conference on the Mathematical Challenges of Big Data, London, December 2016. The author thanks Nicolas Chopin, Petros Dellaportas, Peter Glynn, Emmanuel Gobet, Benjamin Jourdain, and Didier Marteau 
 for helpful conversations. This work was achieved through the Laboratory of Excellence on Financial Regulation (Labex ReFi) under the reference ANR-10-LABX-0095. It benefitted from a French government support managed by the National Research Agency (ANR).

\commentt{}{\bibliographystyle{informs2014}}
\bibliography{poly}
\end{document}